\documentclass[%
 reprint,superscriptaddress,
 amsmath,amssymb,
 aps,
prx,
]{revtex4-1}

\def\I{\mathcal{I}}

\def\r{\rangle}

\def\tr{\mathrm{Tr}}

\usepackage{indentfirst}
\usepackage{float}
\usepackage{graphicx}
\usepackage{epstopdf}
\usepackage{dcolumn}
\usepackage{bm}
\usepackage{xcolor}
\usepackage{color}
\usepackage{url}

\begin{document}


\title{Fluctuation-enhanced quantum metrology}
%
\author{Yu Chen}
\affiliation{Mechanical and Automation Engineering, The Chinese University of HongKong}

\author{Hongzhen Chen}
\affiliation{Mechanical and Automation Engineering, The Chinese University of HongKong}

\author{Jing Liu}
\affiliation{MOE Key Laboratory of Fundamental Physical Quantities Measurement,
Hubei Key Laboratory of Gravitation and Quantum Physics,
PGMF and School of Physics, Huazhong University of Science and Technology,
Wuhan 430074, China }

\author{Zibo Miao}
\affiliation{Mechanical and Automation Engineering, The Chinese University of HongKong}

\author{Haidong Yuan}%
\email{hdyuan@mae.cuhk.edu.hk}
\affiliation{Mechanical and Automation Engineering, The Chinese University of HongKong}





\date{\today}

\begin{abstract}
The main obstacle for practical quantum technology is the noise, which can induce the decoherence and destroy the potential quantum advantages. The fluctuation of a field, which induces the dephasing on the system, is one of the most common noises and widely regarded as detrimental to quantum technologies. Here we show, contrary to the conventional belief, the fluctuation can be used to improve the precision limits in quantum metrology for the estimation of various parameters. Specifically, we show that for the estimation of the direction and rotating frequency of a field, the achieved precisions at the presence of the fluctuation can even surpass the highest precision achievable under the unitary dynamics which have been widely taken as the ultimate limit. We provide explicit protocols, which employs the adaptive quantum error correction, to achieve the higher precision limits with the fluctuating fields. Our study provides a completely new perspective on the role of the noises in quantum metrology. It also opens the door for higher precisions beyond the limit that has been believed to be ultimate.
\end{abstract}

\maketitle


\section{Introduction}
\noindent High precision measurement is one of the major driving forces for technology and science. Quantum metrology, which makes use of quantum mechanical effects to improve the precision limit of parameter estimation\cite{giovannetti2011advances,giovannetti2006quantum,anisimov2010quantum,braunstein1996generalized,paris2009quantum,Fujiwara2008,escher2012general,demkowicz2014using,demkowicz2012elusive,schnabel2010quantum,huelga1997improvement,chin2012quantum,HallPRX,Berry2015,Alipour2014,Beau2017,Liu_2019}, has gained increasing attention in a wide range of applications, such as gravitational wave detection\cite{schnabel2010quantum,ligo2011gravitational}, quantum phase estimation\cite{escher2012general,joo2011quantum,anisimov2010quantum,higgins2007entanglement}, quantum imaging\cite{kolobov1999spatial,lugiato2002quantum,morris2015imaging,roga2016security,tsang2016quantum}, quantum target-detection\cite{shapiro2009quantum,lopaeva2013experimental}, quantum gyroscope\cite{dowling1998correlated} and atomic clock synchronization\cite{bollinger1996optimal,buvzek1999optimal,leibfried2004toward,roos2007designer,derevianko2011colloquium,ludlow2015optical,borregaard2013near}. In practice, however, to achieve the high precision promised by quantum metrology remains a challenging task. The main obstacle is the noise, which can destroy the potential advantages of almost all quantum technologies. The detrimental effect of the noise plays a particularly prominent role in quantum metrology\cite{Fujiwara2008,escher2012general,demkowicz2012elusive,yuanfd}. As to achieve the high precision, the probe needs to be sensitive to the parameter, which typically also makes it vulnerable to noises. Current effort to circumvent this dilemma is typically trying to suppress the noise and restoring the dynamics to unitary evolutions \cite{Plenio2000,Dur2014,Arrad2014,Kessler2014,Ozeri2013,Unden2016,Sekatski2017,Rafal2017,Zhou2018,Layden2018,Layden2019,Zhou2019, Layden2020} as it is believed the unitary evolution leads to the highest precision. 

The fluctuation of a field, which can induce the dephasing on the system, is one of the most common noises in quantum dynamics and widely regarded as detrimental in quantum metrology. Contrary to this belief, we show that while the fluctuation may be harmful for the estimation of the magnitude of a field, it can actually be helpful for the estimation of various other parameters, including the parameters that represent the direction of a field and the frequency of a rotating field. In particular we show that the precision achievable at the presence of the fluctuation can even surpass the highest precision achievable under the ideal unitary dynamics, which has been widely believed as the ultimate limit. We also provide an analysis for general dynamics where the noise and the Hamiltonian can be correlated and present a general formula for the precision achievable with such correlated parametrization. Our study provides a path in quantum metrology that can lead to the higher precision limits beyond what believed to be possible. It can have wide implications in practical applications, such as quantum gyroscope, quantum reference frame alignment etc, where fluctuations are in general unavoidable.

The article is organized as following. In Sec. \ref{sec:model} we give a brief introduction of the essential tools in quantum estimation and introduce the model of a spin interacting with a magnetic field. In Sec.\ref{sec:unitary} we study the estimation of the direction of the field and derive the highest precision achievable under the unitary dynamics. In Sec.\ref{sec:direction} we study the precision that can be achieved at the presence of the fluctuation and show that the precision  outperforms the highest value obtained under the unitary dynamics. We then show that the fluctuation can also improve the precision of estimating the frequency of a rotating field in Sec.\ref{sec:freq}. The optimal measurement and numerical simulations are provided in Sec.\ref{sec:measurement} and Sec.\ref{simulation} respectively. In Sec.\ref{sec:general} we study the general Markovian dynamics where the noise and the Hamiltonian are correlated and provide an analytical analysis on the precision achievable in the asymptotic limit. Sec.\ref{sec:discussion} concludes.

\section{General quantum parameter estimation}
\label{sec:model}
\noindent In this article, we focus on single-parameter estimation where the precision limit of estimating a parameter, $x$, encoded in a quantum state, $\rho_x$, can be calibrated by the quantum Cram\'{e}r-Rao bound(QCRB)\cite{Holevo,helstrom1976quantum,braunstein1994statistical} as $\delta \hat{x} \ge \frac{1}{\sqrt{nJ_Q^x}}$ with $\delta \hat{x}=\sqrt{E[(\hat{x}-x)^2]}$ being the standard deviation of an unbiased estimator($\hat{x}$) and $n$ being the number of copies of the state, here $J_Q^x = \tr[\rho_xL_s^2]$ denotes the quantum Fisher information(QFI) for the parameter $x$(we will use $J_Q^x$ to denote the QFI and $J_C^x$ to denote the classical Fisher information), $L_s$ is the symmetric logarithmic derivative (SLD) which can be obtained from the equation $\frac{\partial \rho_x}{\partial x} = \frac{L_s\rho_x+\rho_xL_s}{2}$. The QFI is closely related to the distance between two neighboring quantum states $\rho_{x}$ and $\rho_{x+dx}$ as\cite{braunstein1994statistical}
\begin{eqnarray}
\label{eq:Bures}
d^2_{Bures}[\rho(x),\rho(x+dx)]=\frac{1}{4}J_Q^xdx^2,
\end{eqnarray}
where $d_{Bures}[\rho_x,\rho_{x+dx}]=\sqrt{2-2F(\rho_x,\rho_{x+dx})}$ is the Bures distance between $\rho_x$ and $\rho_{x+dx}$, where $F(\rho_1,\rho_2)=\tr \sqrt{\rho_1^{\frac{1}{2}}\rho_2\rho_1^{\frac{1}{2}}}$ is the fidelity of two quantum states. Intuitively the faster the state changes with the parameter, the higher the precision. Under the unitary dynamics, the state evolves as $|\psi(t)\rangle=U(t)|\psi(0)\rangle$ with $\frac{\partial U(t)}{\partial t}=-iH(x)U(t)$. In this case the QFI can be computed as
\begin{equation}
J_Q^x=4\Delta h_x^2,
\end{equation}
where $h_x=\int_0^tU^\dagger(\tau)\frac{\partial H(x)}{\partial x} U(\tau)d\tau$\cite{Wilcox1967,Brody_2013,Pang2014PRA,Liu2015} and
\begin{eqnarray}\label{eq:QFI}
\aligned
\Delta h_x^2=\langle\psi(0)|h_x^2|\psi(0)\rangle-\langle\psi(0)|h_x|\psi(0)\rangle^2
\endaligned
\end{eqnarray}
is the variance of $h_x$ with the initial probe state $|\psi(0)\rangle$. 

\begin{figure}[!htp]
\centering
\includegraphics[width=0.5\textwidth]{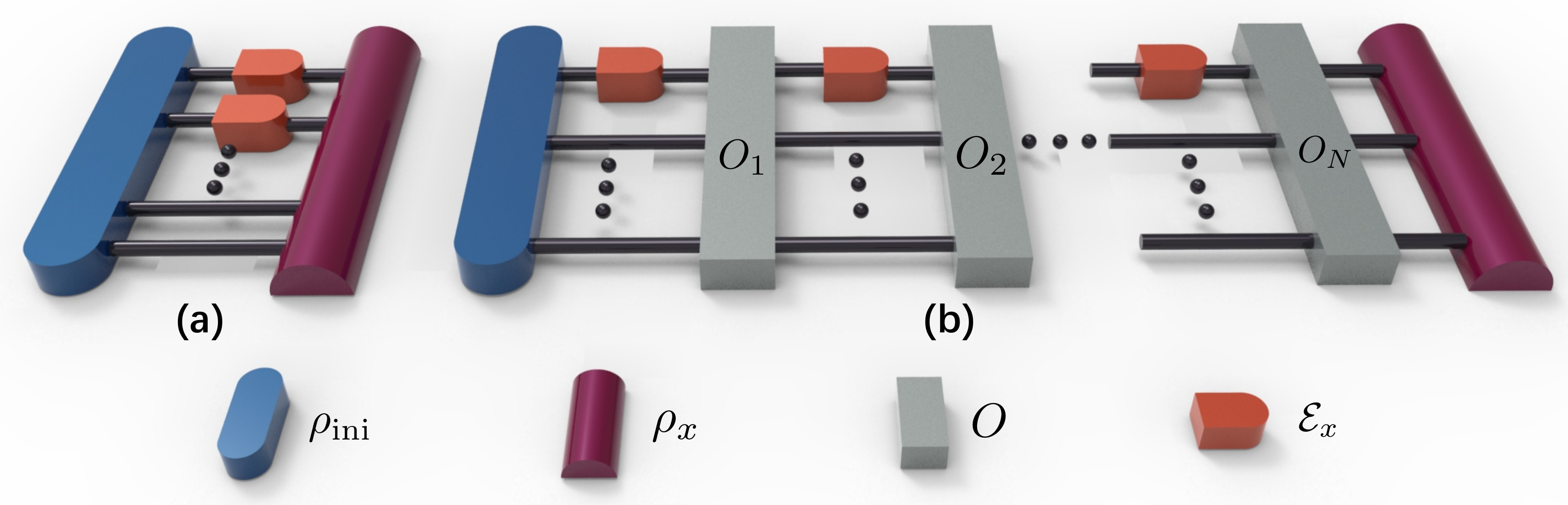}
\caption{\textbf{Schemes for quantum metrology}:(a)parallel scheme; (b)sequential scheme. Here $\rho_{ini}$ denotes the initial probe state, $\mathcal{E}_x$ denotes the dynamics, $O_i$ denotes the control operations and $\rho_x$ denotes the output state. The sequential scheme is the most general scheme which includes the parallel scheme as a special case when taking $O_i$ as the SWAP operations between the system and the $i$th ancilla.}
\label{fig:scheme}
\end{figure}

We consider the general sequential scheme for quantum metrology, as illustrated in Fig.\ref{fig:scheme}(b), which allows adaptive operations during the evolution. The widely used parallel scheme, as shown in Fig.\ref{fig:scheme}(a), is a special case of the general sequential scheme with the control operations taken as SWAP operations between the system and different ancillas. 

We consider using spins to measure a fluctuating magnetic field where the dynamics can be described as
\begin{equation}
\frac{d|\psi\rangle}{dt}=-i[B+\xi(t)]\sigma_{\vec{n}}|\psi\rangle,
\end{equation}
here $\sigma_{\vec{n}}=\vec{n}\cdot \vec{\sigma}$ with $\vec{n}$ denotes the direction of the field with respect to pre-fixed axes, $\vec{\sigma}=(\sigma_1,\sigma_2,\sigma_3)$ is the vector of Pauli matrices, $\xi(t)$ represents the fluctuation which can either arise from the fluctuation of the magnetic field itself or from the environment. We assume that the fluctuation is Markovian with a white spectrum, $E[\xi(t)]=0$, $E[\xi(t)\xi(\tau)]=\gamma\delta(t-\tau)$, which has the most detrimental effects.
For simplicity, we consider a magnetic field in the $XZ$-plane, in which case $\sigma_{\vec{n}(\theta)}=\cos \theta\sigma_1+\sin \theta\sigma_3$ with $\theta$ representing the direction of the magnetic field in the $XZ$-plane.
The estimations of $\theta$ corresponds to the estimation of the direction of the field. It can also corresponds to the orientation of an object in a plane. For example, by attaching a field source (with a known strength) to the object, we can then infer the orientation of the object from the estimation of $\theta$. We are going to show that the estimation of $\theta$ has very different behavior at the presence of the fluctuation, compared with the estimation of $B$.

\section{Precision under the unitary dynamics}
\label{sec:unitary}
\noindent We first consider the precision limits under the unitary dynamics where $\xi(t)=0$. This will be used as benchmarks for the precisions achieved at the presence of the fluctuation. For the estimation of $B$, we have
\begin{eqnarray}
\aligned
h_B&=\int_0^tU^\dagger(\tau)\sigma_{\vec{n}(\theta)}U(\tau)d\tau\\
&=t\sigma_{\vec{n}(\theta)},
\endaligned
\end{eqnarray}
where we have used the fact that $U(\tau)=e^{-i\tau B\sigma_{\vec{n}(\theta)}}$ and $\sigma_{\vec{n}(\theta)}$ commute with each other. The QFI, $J_Q^B=4\Delta h_B^2$, achieves the maximal value $4t^2$ with the optimal initial state $|\psi(0)\rangle=\frac{|\lambda_{\max}\rangle+|\lambda_{\min}\rangle}{\sqrt{2}}$, where $|\lambda_{\max/\min}\rangle$ is the eigenvector of $h_B$ corresponding to the maximal/minimal eigenvalue. If an ancillary spin is available (the free evolution on the ancillary system is null, i.e., Identity), this maximal QFI can also be achieved with the maximally entangled state, $|\psi(0)\rangle=\frac{|00\rangle+|11\rangle}{\sqrt{2}}$. This is the best one can achieve even arbitrary controls are allowed on the system and the ancilla during the evolution\cite{giovannetti2006quantum}.

For the estimation of $\theta$, under the free evolution we have
\begin{eqnarray}
\aligned
h_{\theta}&=\int_0^tU^\dagger(\tau)\frac{\partial H}{\partial \theta}U(\tau)d\tau\\
&=\sin(Bt)\vec{n}_{\theta}\cdot \vec{\sigma},\\
\endaligned
\end{eqnarray}
where $\vec{n}_{\theta}=(-\cos Bt\sin\theta,-\sin Bt,\cos Bt\cos\theta)$. The maximal QFI is then $J_Q^{\theta}=4\sin^2(Bt)$\cite{Pang2014PRA,Jing2015}. Similarly the optimal state can be taken as $|\psi(0)\rangle=\frac{|\lambda_{\max}\rangle+|\lambda_{\min}\rangle}{\sqrt{2}}$ with $|\lambda_{\max/\min}\rangle$ as the eigenvector of $h_{\theta}$. This state depends on $\theta$ thus can only be prepared adaptively according to $h_{\hat{\theta}}$ with $\hat{\theta}$ as the estimated value obtained from accumulated data. But if an ancillary spin is available, the optimal probe state can be taken as $|\psi(0)\rangle=\frac{|00\rangle+|11\rangle}{\sqrt{2}}$, no adaptation is then needed. We note that as $U(\tau)$ and $\frac{\partial H}{\partial \theta}$ do not commute with each other, $h_{\theta}$ does not increase linearly with $t$ which makes $J_Q^{\theta}$ smaller. However, if controls can be employed, proper controls can make $U(\tau)$ commute with $\frac{\partial H}{\partial \theta}$. One choice of such control is to reverse the dynamics and make $U(\tau)=I$\cite{yuan2015optimal,Wiebe2014}. Under such control $h_{\theta}=t\frac{\partial H}{\partial \theta}=Bt(-\sin\theta\sigma_1+\cos\theta\sigma_3)$, the maximal variance of $h_{\theta}$ can reach $B^2t^2$. Such control typically depends on $\theta$ and can only be implemented adaptively, for example, it can be realized by adding a control Hamiltonian $H_c=-B\sigma_{\vec{n}(\hat{\theta})}$ with $\hat{\theta}$ as the estimated value, which converges to the optimal control in the asymptotic limit(see Supplemental Material). The highest QFI for the estimation of $\theta$ achievable under the unitary dynamics with the optimal control strategy is then $J_Q^{\theta}=4B^2t^2$, which can be achieved  when $\hat{\theta}\rightarrow \theta$ in the asymptotic limit.

\section{Precision with the fluctuating field}
\label{sec:direction}
\noindent We now show how the fluctuation affects the precisions of the estimations. We first consider the estimation of $B$ at the presence of the fluctuation. If we take one realization of the fluctuation(a quantum trajectory), the evolution of this trajectory is given by $U(\tau)=e^{-i[B\tau+\int_0^\tau \xi(t)dt]\sigma_{\vec{n}(\theta)}}$, which still commutes with $\frac{\partial H}{\partial B}=\sigma_{\vec{n}(\theta)}$. Thus along one trajectory, $h_B=t\sigma_{\vec{n}(\theta)}$ remains the same and the QFI is bounded above by $4t^2$. The QFI, however, is a statistical quantity which is only meaningful with sufficient repetitions. As for different repetitions we have different realization of the fluctuation, the precision typically decreases resulting in a QFI smaller than $4t^2$(see Supplemental Material). However, if we consider the estimation of $\theta$,  we have $\frac{\partial H}{\partial \theta}=[B+\xi(\tau)](-\sin\theta\sigma_1+\cos\theta\sigma_3)$, thus for a single trajectory,  $h_{\theta}=\int_0^tU^\dagger(\tau)[B+\xi(\tau)](-\sin\theta\sigma_1+\cos\theta\sigma_3)U(\tau)d\tau$. If we can use controls to make $U(\tau)=I$, as the optimal control did in the unitary case, then $h_{\theta}=[Bt+\int_0^t\xi(\tau)d\tau](-\sin\theta\sigma_1+\cos\theta\sigma_3)$. The average variance of $h_{\theta}$ can then reach $E[(Bt+\int_0^t\xi(\tau)d\tau)^2]=B^2t^2+\gamma t$. This is not only beyond what is believed to be achievable under the dephasing dynamics but even surpass the highest value achievable under the unitary dynamics. The noisy dynamics can thus potentially outperform the unitary dynamics. The question is whether this potential can be actually realized. Can we construct explicit protocols to reap this improvement?

At first glance, this does not seem possible. As to design controls that make $U(\tau)=I$, we need to track each trajectory precisely in order to reverse it. This requires a precise knowledge of the fluctuation along each trajectory, which is typically beyond the reach. 
Surprisingly we show that this improvement actually can be achieved. Before we present an explicit control protocol, we first examine the precision that can be achieved under the free evolution at the presence of the fluctuation, i.e., we first compute the precision achievable under the sequential scheme  without adding any control operations.

\begin{figure}[!htp]
\centering
\includegraphics[width=0.5\textwidth]{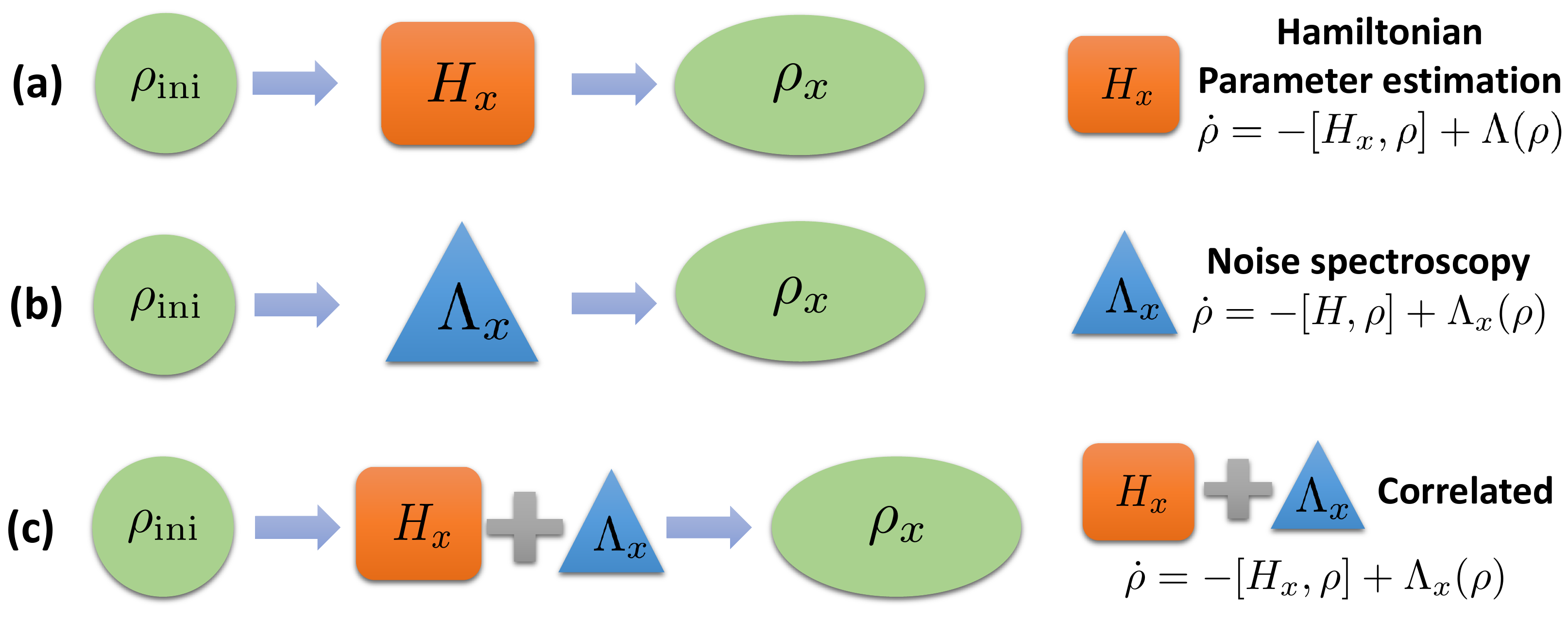}
\caption{\textbf{Parametrization.} (a)Hamiltonian parameter estimation.  The dynamics is described by $\frac{d\rho}{dt}=-i[H_x,\rho]+\Lambda(\rho)$, where only the Hamiltonian, depends on the parameter, $\Lambda(\rho)$, which describes the noisy effect, is independent of the parameter; (b)Noise spectroscopy. The dynamics is $\frac{d\rho}{dt}=-i[H,\rho]+\Lambda_x(\rho)$, where the parameter is only encoded in the noise, the Hamiltonian is independent of the parameter; (c)Correlated parametrization. In the dynamics $\frac{d\rho}{dt}=-i[H_x,\rho]+\Lambda_x(\rho)$, both the Hamiltonian and the noise depend on the parameter.}
\label{fig:hybrid}
\end{figure}

The dynamics of the spin at the presence of the fluctuation can be equivalently described by the master equation
\begin{equation}\label{eq:master}
\frac{d \rho}{d t} = -i[B\sigma_{\vec{n}(\theta)},\rho]+\gamma [\sigma_{\vec{n}(\theta)}\rho \sigma_{\vec{n}(\theta)} - \rho].
\end{equation}
From the master equation, we can immediately see the difference between the estimation of $B$ and the estimation of $\theta$, and understand why the fluctuation can not improve the estimation of $B$. As the parameter $B$ is only encoded in the Hamiltonian, the fluctuation can only affects the precision of $B$ negatively. This is also implicitly assumed in many schemes of quantum metrology, which leads to the belief that the unitary dynamics sets the ultimate limit on the achievable precisions. This assumption, however, does not hold in general. In particular we can see that the parameter $\theta$ is encoded in both the Hamiltonian and the noise. In such correlated parametrization both the Hamiltonian and the noise can contribute to the precision, which makes it possible to achieve higher precision beyond what can be reached under the unitary dynamics. As shown in Fig.\ref{fig:hybrid}, such correlated parametrization differs from the Hamiltonian parameter estimation and the noise spectroscopy\cite{Sinitsyn_2016,Norris2016} studied in literature, which either assume the parameter is only in the Hamiltonian or assume the parameter is only in the noise. The correlated parametrization is also different from the environment-assisted quantum metrology studied previously\cite{Goldstein2011,Cappellaro2012} which utilizes additional spins in the environment as part of the system with limited control.

We now compute the precision that can be achieved under the free evolution. By using an ancilla and preparing the initial state as $|\psi(0)\rangle=\frac{|00\rangle+|11\rangle}{\sqrt{2}}$(for example in NV-center the nuclear spin can be taken as the ancilla as its evolution is much slower than the electron spin, the evolution on the nuclear spin thus can be approximated as Identity), it is straightforward to obtain the QFI as(see Sec. II of the Supplemental material for detail)
\begin{equation}
J_{Q}^{\theta}(t)=2[1-e^{-2\gamma t}\cos(2Bt)].
\end{equation}
By contrast, with the maximally entangled state the QFI under the free unitary dynamics, which corresponds to $\gamma=0$, is
\begin{equation}
J_{QU}^{\theta}(t)=2[1-\cos(2Bt)]=4\sin^2Bt.
\end{equation}
Comparing the two QFI, we can see that the fluctuation displays two competing effects. On the one hand, it provides an extra channel for parametrization which can help improving the precision. This is manifested in the short time regime  with $J_{Q}^{\theta}(t)- J_{QU}^{\theta}(t)\approx 4\gamma t>0$ for $t\ll 1$, exactly the improvement predicted from the heuristic argument. On the other hand, the fluctuation mixes the state which can decrease the precision. This is manifested in the long time regime where under the free evolution the noisy dynamics can perform worse than the unitary dynamics. We now design explicit control strategies that can get rid of the negative effect of the fluctuation but maintain its positive contribution, and show that the improvement from the fluctuation can be kept in the long time regime as well.

\begin{figure}[!htp]
\centering
\includegraphics[width=0.5\textwidth]{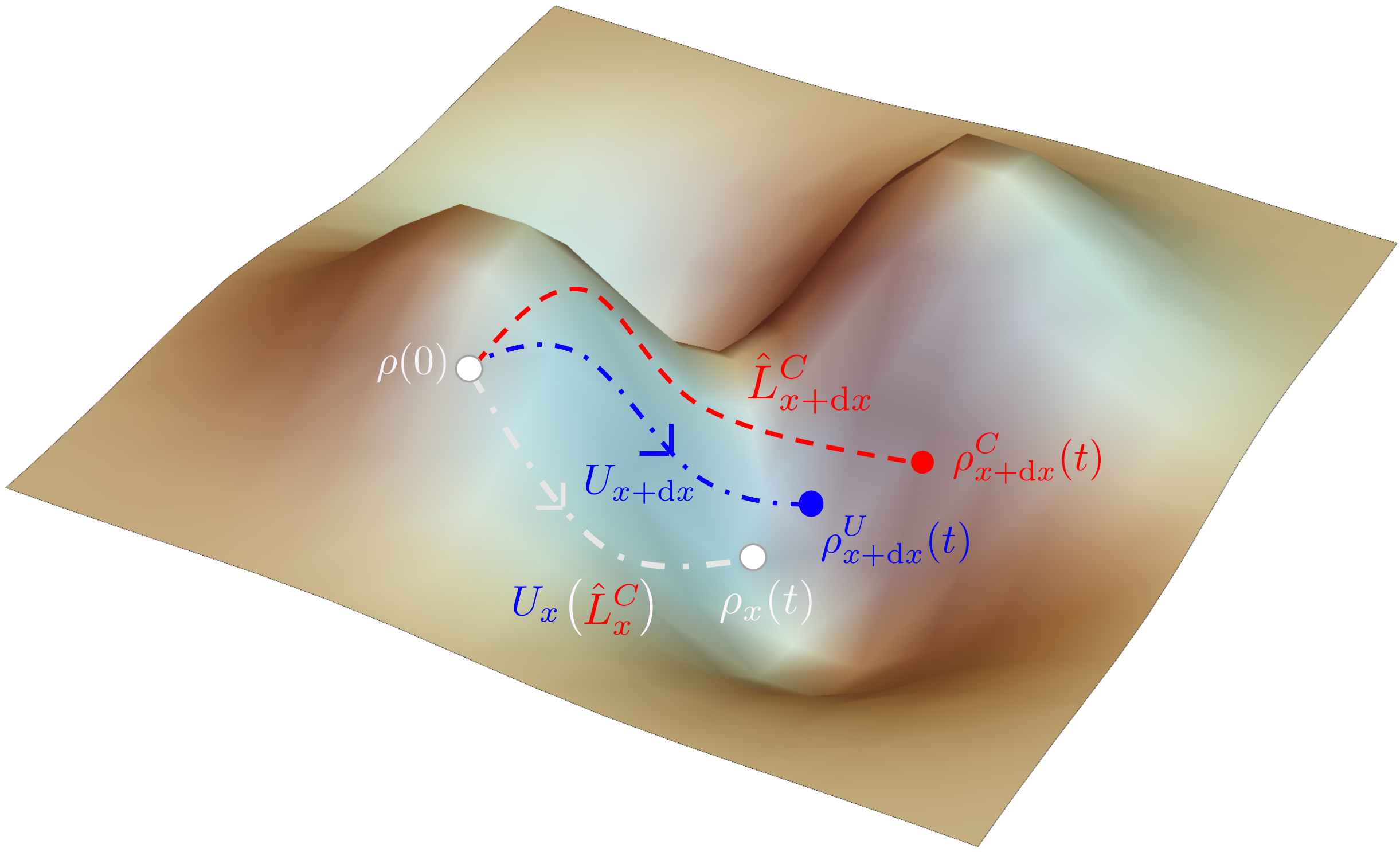}
\caption{\textbf{Illustration of
how the correlated parametrization with the adaptive QEC can lead to more diverging dynamics which improves the precision}. Under the correlated parametrization, the adaptive QEC only corrects the noise at one particular value of the parameter($x$), and at this particular value the corrected dynamics, $L_x^C$, becomes unitary. However, the neighboring dynamics, denoted as $L_{x+dx}^C$, still contains the noise. Compared to the purely unitary dynamics the noise makes $L_{x+dx}^C$ more different from $L_x^C$, which makes the neighboring states further apart (quantified by the Bures distance). This makes the correlated parametrization outperforms the unitary evolution.}
\label{fig:neg}
\end{figure}

We use the quantum error correction(QEC) as the control strategy. At first sight, QEC may seem incompatible with the idea of using the fluctuation to improve the precision. How can the noisy dynamics outperform the unitary ones if the noise is being corrected? The key again lies at the fact that the noise changes with the parameter in the correlated parametrization. In the conventional schemes, the noise is assumed to be independent of the parameter, the QEC corrects the noise regardless of the value of the parameter. After applying the QEC, the dynamics become effectively unitary the noise thus can not contribute to the precision\cite{Plenio2000,Dur2014,Arrad2014,Kessler2014,Ozeri2013,Unden2016,Sekatski2017,Rafal2017,Zhou2018,Layden2018,Layden2019,Zhou2019, Layden2020}. For the correlated parameterization, however, the noise changes with the parameter. A fixed QEC only corrects the noise corresponds to a specific value of the parameter. When the parameter changes, the QEC needs to be adjusted, i.e., the QEC needs to be designed according to the specific value of the parameter and such adaptive QEC only corrects the noise completely at the specific value of the parameter. The dynamics remains noisy when the parameter takes other values. From Eq.(\ref{eq:Bures}), it can be seen that the QFI is determined by the distance between the neighboring states, evolved under the neighboring dynamics. If we denote the noisy dynamics as $\frac{d\rho}{dt}=L_x(\rho)$ with $L_x$ as the super-operator that governs the dynamics, the neighboring dynamics is then $\frac{d\rho}{dt}=L_{x+dx}(\rho)$. Under the correlated parametrization the noises in $L_x$ and $L_{x+dx}$ are different since the noises change with the parameter. The adaptive QEC corrects the noise in $L_x$ which makes the corrected dynamics at $x$, denoted as $\frac{d\rho}{dt}=L_x^C(\rho)$ (here $C$ means with the adaptive QEC), effectively unitary. However, it does not correct the noise in $L_{x+dx}$ completely, i.e., the dynamics with QEC at $x+dx$, denoted as $\frac{d\rho}{dt}=L_{x+dx}^C(\rho)$ (note that $L_{x+dx}^C$ is obtained under the same QEC designed according to the estimated value $\hat{x}$), is still noisy and the noise in it can make the dynamics more different from $L_x^C$ than a purely unitary evolution. Intuitively, the adaptive QEC corrects the part of the noise in $L_{x+dx}$ that is common to the noise in $L_x$, but keeps the different part of the noise in $L_{x+dx}$. By eliminating the common part of the noises, the adaptive QEC reduces the negative effect of the noise, as common noises decreases the distance between the neighboring states. And by keeping the difference of the noise, the adaptive QEC preserves the positive effect of the noise, as the difference of the noise can make the neighboring dynamics more divergent.  Take the qubits for example, the fidelity between the neighboring states can be decomposed into two parts as\cite{HUBNER1992239}
\begin{equation}
\label{eq:fidelityqubit}
F(\rho_x,\rho_{x+dx})=Tr(\rho_x\rho_{x+dx})+2\sqrt{det(\rho_x)det(\rho_{x+dx})}.
\end{equation}
In the correlated parametrization, different noises in the neighboring dynamics make $\rho_x$ and $\rho_{x+dx}$ more different which has the effect of reducing the first term thus increase the Bures distance $d_{Bures}^2=2-2F(\rho_x,\rho_{x+dx})$. However, if there are no control strategies, the states will become mixed and the second term in Eq.(\ref{eq:fidelityqubit}) increases which can override the positive contributions of the noise. By applying the adaptive QEC which corrects the common part of the noises in $\rho_x$ and $\rho_{x+dx}$, we are able to keep the positive contribution of the noise while reducing the negative effect. In the asymptotic limit when the estimated value $\hat{x}\rightarrow x$, under the adaptive QEC that is designed according to $\hat{x}$, $\rho_x$ will become a pure state, the second term in Eq.(\ref{eq:fidelityqubit}) vanishes, i.e., the negative effect is eliminated. But the positive contribution remains as the noise in $\rho_{x+dx}$ is not completely eliminated. Only the part that is common with the noise in $\rho_x$ is eliminated but the part that is different remains which helps reduce the first term in Eq.(\ref{eq:fidelityqubit}) as it makes $\rho_{x+dx}$ more different from $\rho_x$ (similar effects holds in higher dimensional space although a general formula to decompose the fidelity this way has not been known). Thus the noise in the correlated parametrization can make the neighboring states further apart than the states from purely unitary dynamics, as illustrated in Fig.\ref{fig:neg}. We emphasize that the positive contribution comes from the fact that the noise changes with the parameter in the correlated parametrization, the adaptive QEC is just a tool to keep the positive contribution of the noise from being smeared by the negative mixing effect. The adaptive QEC itself is not the source of the outperformance over the unitary dynamics. Without the correlated parametrization, the QEC can at most recover the performance of the unitary evolution as in previous studies with QEC\cite{Plenio2000,Dur2014,Arrad2014,Kessler2014,Ozeri2013,Unden2016,Sekatski2017,Rafal2017,Zhou2018,Layden2018,Layden2019,Zhou2019, Layden2020}.

%

We now provide an explicit protocol with the adaptive QEC. With an acillary spin (unless specified, an operator, $A$, will be understood as $A\otimes I$ with $A$ on the probe spin and Identity on the acillary spin), we choose the code space as $\{|C_0\rangle=|+_{\hat{\theta}}\rangle|+_{\hat{\theta}}\rangle, |C_1\rangle=|-_{\hat{\theta}}\rangle|-_{\hat{\theta}}\rangle\}$, where $|+_{\hat{\theta}}\rangle=-\cos\frac{\hat{\theta}}{2}|0\rangle + \sin\frac{\hat{\theta}}{2}|1\rangle$, $|-_{\hat{\theta}}\rangle=\sin\frac{\hat{\theta}}{2}|0\rangle + \cos\frac{\hat{\theta}}{2}|1\rangle$ are the eigenvectors of $\frac{\partial H(\hat{\theta})}{\partial \theta}=B(-\sin \hat{\theta}\sigma_1+\cos \hat{\theta}\sigma_3)$. Here $\hat{\theta}$ is the estimation of $\theta$ obtained from the prior knowledge and the accumulated measurement data. Together with $|C_2\rangle=|-_{\hat{\theta}}\rangle|+_{\hat{\theta}}\rangle$ and $|C_3\rangle=|+_{\hat{\theta}}\rangle|-_{\hat{\theta}}\rangle$, they form a basis for the space of two spins. It is easy to check that $\sigma_{\vec{n}(\hat{\theta})}|C_0\rangle=|C_2\rangle$ and $\sigma_{\vec{n}(\hat{\theta})}|C_1\rangle=|C_3\rangle$. If we prepare the initial probe state as $\frac{|C_0\rangle+|C_1\rangle}{\sqrt{2}}$, which is just the maximally entangled state $\frac{|00\rangle+|11\rangle}{\sqrt{2}}$, both the Hamiltonian and the noisy operator will drive the state out of the code space, however, by applying the QEC, which steers the state back to the code space, we can make the evolution as the Identity in the asymptotic limit when $\hat{\theta}$ converges to $\theta$.

\begin{figure}[!htp]
\centering
\includegraphics[width=0.5\textwidth]{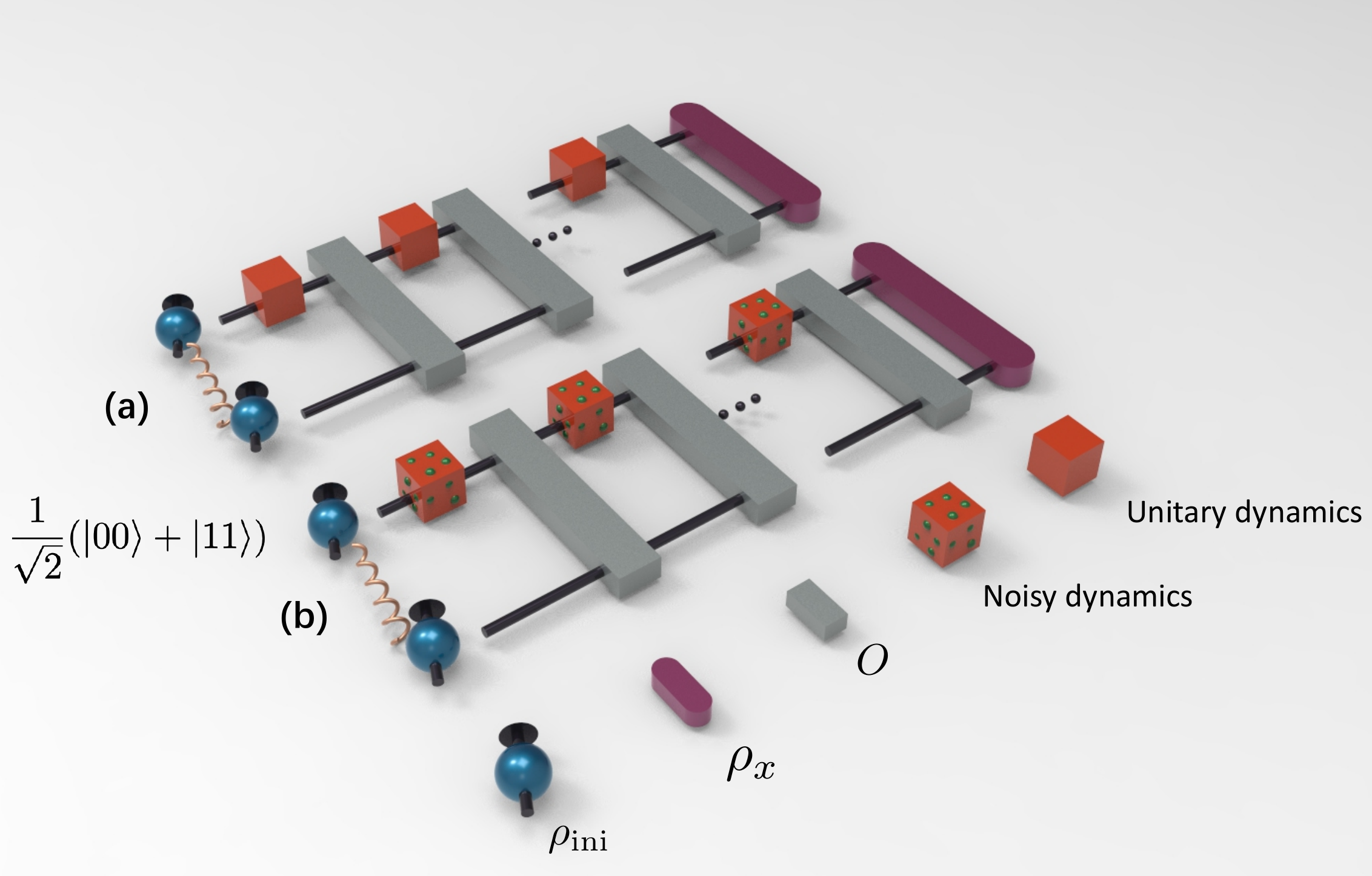}
\caption{\textbf{Scheme for the estimation of $\theta$ under the unitary and noisy dynamics.} The initial probe state is the maximally entangled state for both cases. With the optimal controls, the maximal QFI achievable under the unitary dynamics is $4B^2t^2$, which is believed to be the ultimate limit. However, under the noisy dynamics with the fluctuation the QFI surpasses this limit.} \label{fig:compare}
\end{figure}

Specifically, suppose $\rho_C(0)$ is in the code space, then after a small period $dt$, the state evolves to $\rho(dt)=$
\begin{equation}
\rho_C(0)-i[B\sigma_{\vec{n}(\theta)},\rho_C(0)]dt+\gamma (\sigma_{\vec{n}(\theta)}\rho_C(0) \sigma_{\vec{n}(\theta)} - \rho_C(0))dt,
\end{equation}
which can be out of the code space.
A recovery operation, which consists of two Kraus operators, $\{\Pi_C,\Pi_C\sigma_{\vec{n}(\hat{\theta})}\}$, is then applied on the state, here $\Pi_C=|C_0\rangle\langle C_0|+|C_1\rangle\langle C_1|$ and $\sigma_{\vec{n}(\hat{\theta})}=\cos\hat{\theta}\sigma_1+\sin\hat{\theta}\sigma_3$. For any $\rho_C(0)$ in the code space, we have $\Pi_C\rho_C(0)\Pi_C=\rho_C(0)$ and $\Pi_C\sigma_{\vec{n}(\hat{\theta})}\rho_C(0)=\rho_C(0)\sigma_{\vec{n}(\hat{\theta})}\Pi_C=0$. The state after the recovery operation can then be obtained as
\begin{eqnarray}
\aligned
&\rho_C(dt)\\
=&\Pi_C\rho(dt)\Pi_C+\Pi_C\sigma_{\vec{n}(\hat{\theta})}\rho(dt)\sigma_{\vec{n}(\hat{\theta})}\Pi_C\\
=&\rho_C(0)-i[B\Pi_C\sigma_{\vec{n}(\theta)}\Pi_C,\rho_C(0)]\\
&+\gamma[\Pi_C\sigma_{\vec{n}(\theta)}\Pi_C\rho_C(0)\Pi_C\sigma_{\vec{n}(\theta)}\Pi_C\\
&+\Pi_C\sigma_{\vec{n}(\hat{\theta})}\sigma_{\vec{n}(\theta)}\Pi_C\rho_C(0)\Pi_C\sigma_{\vec{n}(\theta)}\sigma_{\vec{n}(\hat{\theta})}\Pi_C-\rho_C(0)]dt,
\endaligned
\end{eqnarray}
 where $\Pi_C\sigma_{\vec{n}(\theta)}\Pi_C=\sin(\theta-\hat{\theta})(|C_0\rangle\langle C_0|-|C_1\rangle\langle C_1|)$, $\Pi_C\sigma_{\vec{n}(\hat{\theta})}\sigma_{\vec{n}(\theta)}\Pi_C=\cos(\theta-\hat{\theta})(|C_0\rangle\langle C_0|+|C_1\rangle\langle C_1|)$. Denote $\sigma_3^C=|C_0\rangle\langle C_0|-|C_1\rangle\langle C_1|$, we then obtain the dynamics of the recovered state as
\begin{eqnarray}
\label{eq:dynamicsQEC}
\aligned
\frac{d\rho_C}{dt}=&-i[B\sin(\theta-\hat{\theta})\sigma_3^C,\rho_C]\\
&+\gamma\sin^2(\theta-\hat{\theta})[\sigma_3^C\rho_C\sigma_3^C-\rho_C].
\endaligned
\end{eqnarray}

If the initial state is taken as $\frac{|C_0\rangle+|C_1\rangle}{\sqrt{2}}$, the final state can be easily obtained as $\rho_C(t)=\frac{1}{2}(|C_0\rangle\langle C_0|+|C_1\rangle\langle C_1|+e^{-gt}|C_0\rangle\langle C_1|+e^{-g^*t}|C_1\rangle\langle C_0|)$
with $g=2iB\sin(\theta-\hat{\theta})+2\gamma\sin^2(\theta-\hat{\theta})$. The QFI is $J_Q^{\theta}(t)=$
\begin{equation}\label{eq.rhocqfi}
 \begin{aligned}
    \\
    4t^2\cos^2(\theta-\hat{\theta})\frac{B^2(1-e^{-4\gamma t\sin^2(\theta-\hat{\theta}) })+4\gamma^2\sin^2(\theta-\hat{\theta}) }{e^{4\gamma t\sin^2(\theta-\hat{\theta}) }-1},
  \end{aligned}
\end{equation}
which, expanded to the second order of $d\theta=\hat{\theta}-\theta$, is
\begin{equation}
  \begin{aligned}
    J_Q^{\theta}(t)=&4B^2t^2+4\gamma t \\
    &- 4B^2d\theta^2t^2(1+2\frac{\gamma^2}{B^2}+\frac{\gamma}{B^2t}+4\gamma t)+o(d\theta^4).
  \end{aligned}
  \label{eq:qfi1}
\end{equation}
It is now easy to see that the QFI achieves the maximal value, $4B^2t^2+4\gamma t$, at the asymptotic limit when $d\theta=0$.
This is also exactly the value predicted from the heuristic argument\cite{note2}. In the finite regime where $\hat{\theta}$ is not equal to $\theta$, the $t^2$ scaling is maintained within the time order of $(Bd\theta)^{-1}$. We note that this is not much of a restriction, the evolution time should be restricted to the same order even under the unitary evolution in order to avoid the phase ambiguities(see Supplement Material). As shown in Fig.\ref{fig:finite}(a), in the finite regime under the adaptive QEC, with quite an amount of estimation error the QFI surpasses the highest value achievable under the unitary dynamics. And in this case the stronger the fluctuation, the more the improvement. The highest precision achievable under the unitary dynamics thus does not play the role of a fundamental bound.

\begin{figure}[!htp]
\centering
\includegraphics[width=0.5\textwidth]{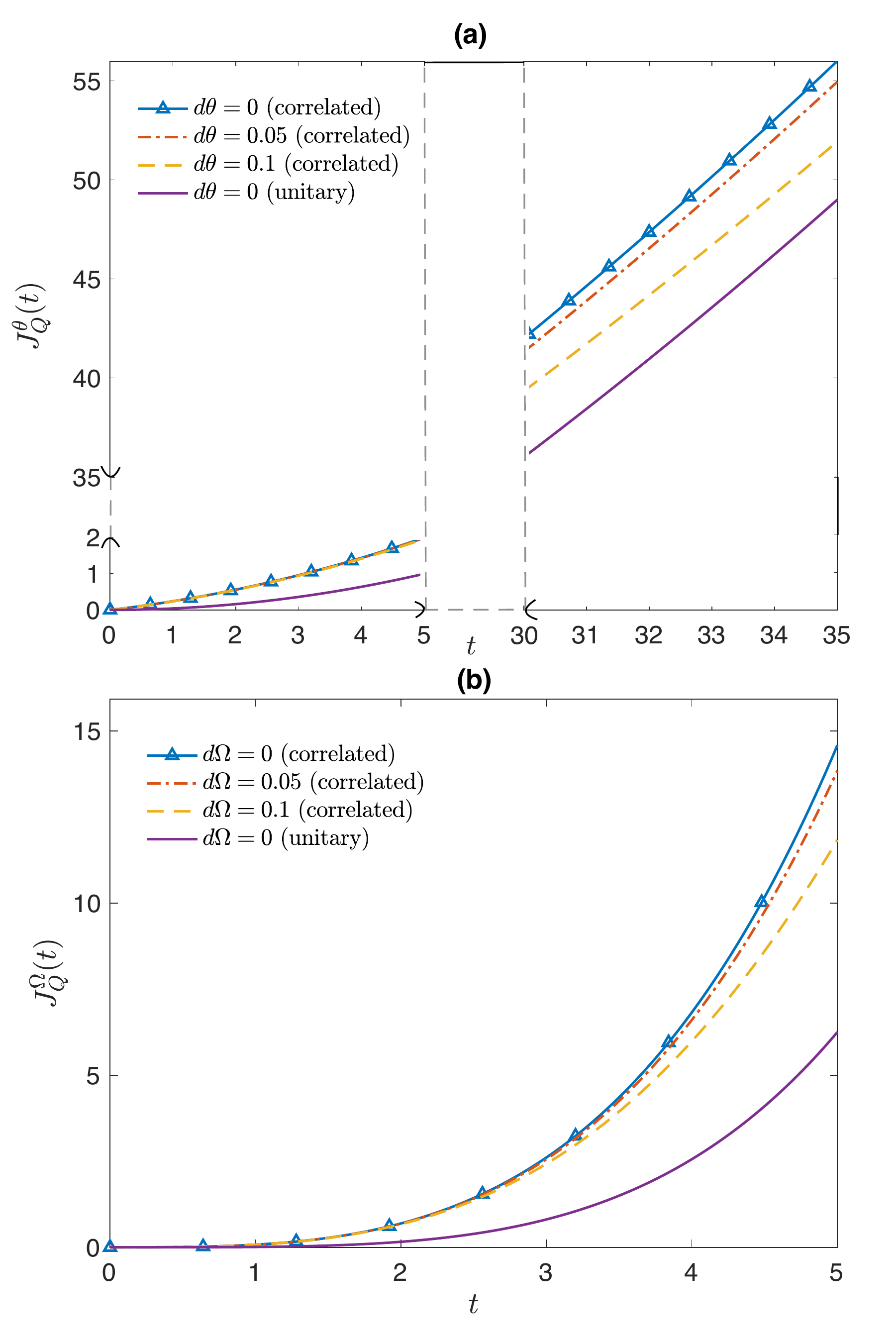}
\caption{\textbf{The QFI for the estimation of $\theta$ (and $\Omega$) under the adaptive QEC}. (a)The QFI for the estimation of $\theta$, where the true value is taken as $\frac{\pi}{4}$. The bottom solid line corresponds to the highest QFI($4B^2t^2$) one can achieve under the optimally controlled unitary dynamics in the asymptotical limit when $d\theta=0$, while the other three lines corresponds to the QFI under the adaptive QEC with $d\theta=\hat{\theta}-\theta$ taking $0$, $0.05$ and $0.1$ respectively. The initial state is the maximally entangled state for all cases. (b) The QFI for the estimation of $\Omega$, where the true value is taken as $0.5$. The bottom solid line corresponds to the highest QFI($B^2t^4$) one can achieve under the optimally controlled unitary dynamics in the asymptotical limit when $d\Omega=0$, while the other three lines corresponds to the QFI under the adaptive QEC with $d\Omega=\hat{\Omega}-\Omega$ taking $0$, $0.05$ and $0.1$ respectively. The initial state is the maximally entangled state for all cases. Here $B$ is taken as $0.1$, $\gamma$ is taken as $0.05$.}
\label{fig:finite}
\end{figure}
\section{Estimation of the rotating frequency of a fluctuating field}\label{sec:freq}

\noindent The fluctuation can also improve the precisions for the estimation of other parameters. For example, it can be used to improve the precision for the estimation of the frequency of a rotating field.

Consider a fluctuating field that rotates in the $XY$-plane with a frequency $\Omega$ and the dynamics of a spin is then described by
\begin{equation}
\frac{d|\psi\rangle}{dt}=-i[B+\xi(t)]\sigma_{\vec{n}}(\Omega, t)|\psi\rangle,
\end{equation}
here $\xi(t)$ is the fluctuation, $\sigma_{\vec{n}}(\Omega,t)=\cos(\Omega t) \sigma_1-\sin(\Omega t) \sigma_2$ with $\Omega$ as the frequency to be estimated. If the fluctuation has a faster time scale than the rotation, the dynamics can be equivalently described by the master equation
\begin{equation}
\label{eq:frequency}
\frac{d\rho}{dt}=-i[B\sigma_{\vec{n}}(\Omega, t), \rho]+\gamma[\sigma_{\vec{n}}(\Omega, t)\rho\sigma_{\vec{n}}(\Omega,t)-\rho].
\end{equation}
The estimation of the frequency of a rotating field is a fundamental problem in quantum metrology and has been extensively studied \cite{Pang2017,Schmitt832,Boss837,Naghiloo2017,Glenn2018}. The highest QFI is believed to be achieved under the optimally controlled unitary dynamics, which takes the value $B^2t^4$ in the asymptotic limit\cite{Pang2017}. This has been widely regarded as the ultimate limit for the estimation of the frequency. We now show that this can be surpassed at the presence of the fluctuation.

Again we first give a heuristic argument. For each trajectory we have $h_{\Omega}=$
\begin{eqnarray}
\nonumber
\aligned
&\int_0^tU^\dagger(\tau)\frac{\partial H}{\partial \Omega}U(\tau)d\tau\\
=&\int_0^tU^\dagger(\tau)\{-\tau[B+\xi(\tau)](\sin\Omega \tau\sigma_1+\cos\Omega \tau\sigma_2)\}U(\tau)d\tau\\
=&\int_0^tU^\dagger(\tau)\{-\tau[B+\xi(\tau)]e^{-i\frac{\Omega}{2}\sigma_3}\sigma_2e^{i\frac{\Omega}{2}\sigma_3}\}U(\tau)d\tau.
\endaligned
\end{eqnarray}
If we can make $U(\tau)=e^{-i\frac{\Omega}{2}\sigma_3}$, then $h_{\Omega}=\int_0^t-\tau[B+\xi(\tau)]\sigma_2d\tau=-[\frac{1}{2}Bt^2+\int_0^t\tau\xi(\tau)d\tau]\sigma_2$, the average variance of $h_{\Omega}$ is then given by $E[(\frac{1}{2}Bt^2+\int_0^t\tau\xi(\tau)d\tau)^2]=\frac{1}{4}B^2t^4+\frac{1}{3}\gamma t^3$. The QFI can thus potentially reach $B^2t^4+\frac{4}{3}\gamma t^3$, which is higher than the highest value achievable under the unitary dynamics. We show explicitly how this can be achieved with the adaptive QEC.


As the noise in this case not only changes with the parameter, but also changes with time, the code space of the adaptive QEC will also be time-dependent. We choose the basis of the code space at time $t$ as $\{|C_0(t)\rangle=|+(\hat{\Omega},t)\rangle|0\rangle, |C_1(t)\rangle=|-(\hat{\Omega},t)\rangle|1\rangle\}$, where $|\pm(\hat{\Omega},t)\rangle$ are the eigenvectors of $h_{\hat{\Omega}}(t)=\frac{\partial H(\hat{\Omega}, t)}{\partial \Omega}=-Bt(\sin\hat{\Omega} t\sigma_1+\cos\hat{\Omega} t\sigma_2)=-Bte^{-i\hat{\Omega}t\sigma_3}\sigma_2e^{i\hat{\Omega}t\sigma_3}$. Here $\hat{\Omega}$ is the estimated value of $\Omega$ obtained from previous data.

Suppose at time $t$ the probe state, denoted as $\rho_C(t)$, is in the code space, then after a period of $dt$, it evolves to
\begin{eqnarray}
\aligned
\rho(t+dt)=&\rho_C(t)-i[B\sigma_{\vec{n}}(\Omega, t), \rho_C(t)]dt\\
&+\gamma[\sigma_{\vec{n}}(\Omega, t)\rho_C(t)\sigma_{\vec{n}}(\Omega,t)-\rho_C(t)]dt.
\endaligned
\end{eqnarray}
We then apply a recovery operation, which consists of two Kraus operators, $\{\Pi_C(t),
\Pi_C(t)\sigma_{\vec{n}}(\hat{\Omega},t)\}$, on the state, where $\Pi_C(t)=|C_0(t)\rangle\langle C_0(t)|+|C_1(t)\rangle\langle C_1(t)|$ and $\sigma_{\vec{n}}(\hat{\Omega},t)=\cos(\hat{\Omega} t) \sigma_1-\sin(\hat{\Omega} t) \sigma_2$. The state after the recovery operation can be obtained as
\begin{eqnarray}
\aligned
\tilde{\rho}_C(t+dt)&=\Pi_C(t)\rho(t+dt)\Pi_C(t)\\
&+\Pi_C(t)\sigma_{\vec{n}}(\hat{\Omega},t)\rho(t+dt)\sigma_{\vec{n}}(\hat{\Omega},t)\Pi_C(t)\\
&=\rho_C(t)-i[B\sin(\Omega-\hat{\Omega})t \sigma_3^C(t),\rho_C(t)]dt\\
&+\gamma\sin^2(\Omega-\hat{\Omega})t[\sigma_3^C(t)\rho_C(t)\sigma_3^C(t)-\rho_C(t)]dt,
\endaligned
\end{eqnarray}
where $\sigma_3^C(t)=|C_0(t)\rangle\langle C_0(t)|-|C_1(t)\rangle\langle C_1(t)|$.
After the recovery operation, we apply another unitary operation, $\tilde{U}(dt)$, on the state. This unitary operation is to rotate the code space at $t$ to the code space at $t+dt$, i.e., $\tilde{U}(dt)\Pi_C(t)\tilde{U}^\dagger(dt)=\Pi_C(t+dt)$. As $|C_0(t)\rangle,|C_1(t)\rangle$ are the eigenvectors of
$h_{\hat{\Omega}}(t)=-Bte^{-i\hat{\Omega}t\sigma_3}\sigma_2e^{i\hat{\Omega}t\sigma_3}$, we have $|C_0(t)\rangle=e^{i\frac{\hat{\Omega}}{2}t\sigma_3}|C_0(0)\rangle$ and $|C_1(t)\rangle=e^{i\frac{\hat{\Omega}}{2}t\sigma_3}|C_1(0)\rangle$, then $\frac{d}{dt}\Pi_C(t)=i[\frac{\hat{\Omega}}{2}\sigma_3, \Pi_C(t)]$, thus $\tilde{U}(dt)=e^{i\frac{\hat{\Omega}}{2}\sigma_3dt}$.
After applying the unitary $\tilde{U}(dt)$, the state becomes
\begin{eqnarray}
\aligned
&\rho_C(t+dt)\\
=&\tilde{\rho}_C(t+dt)+i[\frac{\hat{\Omega}}{2}\sigma_3,\tilde{\rho}_C(t+dt)]dt\\
=&\rho_C(t)+i[\frac{\hat{\Omega}}{2}\sigma_3,\rho_C(t)]dt\\
&-i[B\sin(\Omega-\hat{\Omega})t \sigma_3^C(t),\rho_C(t)]dt\\
&+\gamma\sin^2(\Omega-\hat{\Omega})t[\sigma_3^C(t)\rho_C(t)\sigma_3^C(t)-\rho_C(t)]dt.
\endaligned
\end{eqnarray}
The dynamic of the corrected state can then be obtained as
\begin{eqnarray}
\label{eq:frequencyqec}
\aligned
\frac{d\rho_C(t)}{dt}=&i[\frac{\hat{\Omega}}{2}\sigma_3-B\sin(\Omega-\hat{\Omega})t \sigma_3^C(t),\rho_C(t)]\\
&+\gamma\sin^2(\Omega-\hat{\Omega})t[\sigma_3^C(t)\rho_C(t)\sigma_3^C(t)-\rho_C(t)].
\endaligned
\end{eqnarray}
To compute the final state under this dynamics, we move to the rotating frame with $\rho_R(t)=e^{-i\frac{\hat{\Omega}}{2}\sigma_3t}\rho_C(t)e^{i\frac{\hat{\Omega}}{2}\sigma_3t}$, then
\begin{eqnarray}
\aligned
\frac{d\rho_R(t)}{dt}=&-i[B\sin(\Omega-\hat{\Omega})t \sigma_3^C(0),\rho_R(t)]\\
&+\gamma\sin^2(\Omega-\hat{\Omega})t[\sigma_3^C(0)\rho_R(t)\sigma_3^C(0)-\rho_R(t)],
\endaligned
\end{eqnarray}
where we have used the fact that $e^{-i\frac{\hat{\Omega}}{2}\sigma_3t}\sigma_3^C(t)e^{i\frac{\hat{\Omega}}{2}\sigma_3t}=\sigma_3^C(0)$.
Note that $\rho_R(t)$ has the same QFI as $\rho_C(t)$ since $e^{-i\frac{\hat{\Omega}}{2}\sigma_3t}$ only depends on $\hat{\Omega}$. When the initial state is taken as $\frac{|C_0(0)\rangle+|C_1(0)\rangle}{\sqrt{2}}$, we have
\begin{eqnarray}
\aligned
    \rho_R(t)=\frac{1}{2}(&|C_0(0)\rangle\langle C_0(0)|+|C_1(0)\rangle\langle C_1(0)|\\
    +&e^{-\int_0^tg(\tau)d\tau}|C_0(0)\rangle\langle C_1(0)|\\
    +&e^{-\int_0^tg^*(\tau)d\tau}|C_1(0)\rangle\langle C_0(0)|),
\endaligned
\end{eqnarray}
where $g=2\sin(\Omega-\hat{\Omega})t[iB+\gamma\sin(\Omega-\hat{\Omega})t]$.
Up to the second order of $d\Omega=\hat{\Omega}-\Omega$, the state's QFI is
\begin{eqnarray}
  \begin{aligned}
    J_Q^{\Omega}(t)= &B^2t^4+\frac{4}{3}\gamma t^3\\
    &-(\frac{1}{2}B^2t^3+\frac{4}{5}\gamma t^2+\frac{4}{3}B^2\gamma t^4+\frac{8}{9}\gamma^2 t^3)d\Omega^2t^3, 
  \end{aligned}
\end{eqnarray}
%
which achieves the maximal value $B^2t^4+\frac{4}{3}\gamma t^3$ in the asymptotic limit when $d\Omega\rightarrow 0$, the same as predicted from the heuristic argument. In Fig.\ref{fig:finite}(b) we simulated the exact QFI with some finite $d\Omega$, it can be seen that  with quite an amount of estimation error the QFI surpasses the highest value achievable under the unitary dynamics. 
\section{Optimal measurement}\label{sec:measurement}

\noindent For the estimation of $\theta$, the state at time $t$ is
$\rho_C(t) = \frac{1}{2}(|C_0\rangle\langle C_0|+|C_1\rangle \langle C_1|+e^{-gt}|C_0\rangle \langle C_1| + e^{-g^*t}|C_1\rangle \langle C_0|)$,
where $g = 2iB\sin(\theta-\hat{\theta})+2\gamma \sin^2(\theta-\hat{\theta})$. The optimal measurement is the projective measurement on the eigenvectors of $\sigma_x^C=|C_0\rangle\langle C_1|+|C_1\rangle\langle C_0|$, i.e., the projective measurements on the two states, $\frac{|C_0\rangle+ |C_1\rangle}{\sqrt{2}}$ and $\frac{|C_0\rangle- |C_1\rangle}{\sqrt{2}}$(note that $|C_0\rangle$ and $|C_1\rangle$ only depend on $\hat{\theta}$). In the asymptotic limit when $\hat{\theta}=\theta$, this achieves the highest Fisher information, $4B^2t^2+4\gamma t$. The classical Fisher information under this measurement in the finite regime where $d\theta=\hat{\theta}-\theta\neq 0$, can also be obtained explicitly as
\begin{equation}
  \begin{aligned}
    J_C^{\theta}(t)\approx&4B^2t^2+4\gamma t\\
    &- 4B^2d\theta^2t^2(1+2\frac{\gamma^2}{B^2}+\frac{\gamma}{B^2t}+4\gamma t)+o(d\theta^4).
  \end{aligned}
\end{equation}
This is the same as the QFI up to the second order of $d\theta$, showing that this measurement is optimal.

For the estimation of $\Omega$, the optimal measurement is the projective measurement on the the eigenvectors of $\sigma_x^C(t)=|C_0(t)\rangle\langle C_1(t)|+|C_1(t)\rangle\langle C_0(t)|$, i.e., the projective measurements on the two states, $\frac{|C_0(t)\rangle\pm |C_1(t)\rangle}{\sqrt{2}}$. This achieves the highest QFI, $B^2t^4+\frac{4\gamma t^3}{3}$, in the asymptotic limit. In the finite regime, up to the second order of $d\Omega=\hat{\Omega}-\Omega$ the classical Fisher information under this measurement can be explicitly obtained as
\begin{eqnarray}
\aligned
J_C^{\Omega}& =B^2 t^4+\frac{4 \gamma  t^3}{3}\\
-&(\frac{1}{2}B^2t^3+\frac{4}{5}\gamma t^2+\frac{4}{3}B^2\gamma t^4+\frac{8}{9}\gamma^2 t^3)d\Omega^2t^3+o\left(d\Omega^3\right),
\endaligned
\end{eqnarray}
which is the same as the QFI up to the second order of $d\Omega$.

\section{Simulation}\label{simulation}
We now demonstrate the protocol with numerical simulations. For the estimation of $\theta$ the simulation consists of the following steps:
\begin{enumerate}
  \item Make an initial estimation of $\theta$, denoted as $\hat{\theta}_0$. This can be obtained from a prior knowledge or just a guess;
  \item Design a QEC code, $\{|C_0\rangle, |C_1\rangle\}$, according to the estimated value of $\theta$,
  \item Prepare the initial state as $|\Psi\rangle=\frac{|C_0\rangle+|C_1\rangle}{\sqrt{2}}$ and let it evolve under the dynamics with the QEC, then perform the projective measurement on $\frac{|C_0\rangle\pm|C_1\rangle}{\sqrt{2}}$. This leads to the measurement result with the probability distribution
    $p(+|\theta,\hat{\theta})
    =\frac{1}{2}\left( 1+e^{-2t\gamma\sin^2(\theta-\hat{\theta})}\cos[2Bt\sin(\theta-\hat{\theta})] \right)$,
    $p(-|\theta,\hat{\theta})=\frac{1}{2}\left( 1-e^{-2t\gamma\sin^2(\theta-\hat{\theta})}\cos[2Bt\sin(\theta-\hat{\theta})] \right)$.
 \item Repeat step 3 for $m$ times($m$ is taken as $10$ in our simulation) and record the measurement results as $\bm{x}^0=\{x_1,...,x_m\}$ with $x_i\in\{+,-\}$.
 \item Use the maximal likelihood to update the estimator as $\hat{\theta}_1=\mathrm{argmax}_{\theta}L^m[x|\theta,\hat{\theta}_0]$ where $L^m[x^0|\theta,\hat{\theta}_0]=p[x_1|\theta,\hat{\theta}_0)p[x_2|\theta,\hat{\theta}_0]\cdots p[x_m|\theta,\hat{\theta}_0]$.
  \item Based on $\hat{\theta}_1$, repeat step 2,3 and record the measurement results as $\bm{x}^1$ and update the estimation with all previous data as
  $\hat{\theta}_2=\mathrm{argmax}_{\theta}L^m[x^1|\theta,\hat{\theta}_1]L^m[x^0|\theta,\hat{\theta}_0]$. Repeat it for $K$ times to get $\hat{\theta}_K=\mathrm{argmax}_{\theta} \Pi_{l=0}^{K-1} L^m(\theta,\hat{\theta}_l)$.
\end{enumerate}

\begin{figure}
  \includegraphics[width=0.55\textwidth]{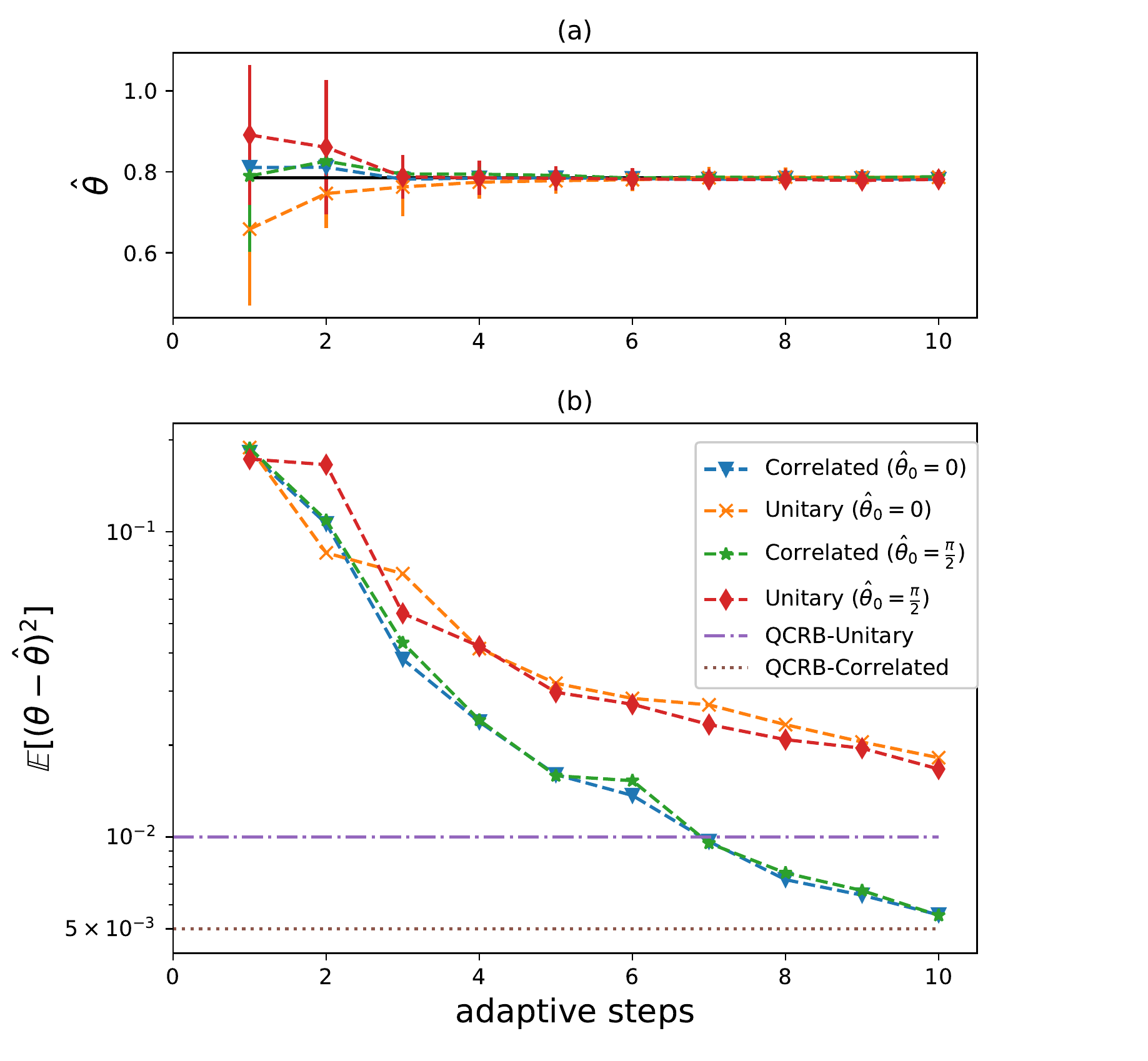}
  \caption{\textbf{Simulation for the estimation of $\theta$.} (a) Simulation for the update of the estimator $\hat{\theta}$ with the initial guess as $0$ and $\frac{\pi}{2}$ respectively, where the true value is taken as $\theta=\frac{\pi}{4}$.  (b)Simulation of the mean squared error under the adaptive QEC by repeating the estimation for 1000 times. The updating process for the adaptive control of the unitary dynamics is also plotted for comparison. The highest precision determined by the QCRB, $\delta \hat{\theta}^2\geq \frac{1}{mKJ_Q^{\theta}}$, is also marked, where $mK=100$ is the number of the measurements to obtain $\hat{\theta}_K$, $J_Q^{\theta}$ is taken as the maximal QFI at $d\theta=0$, i.e., $J_Q^{\theta}=4B^2t^2$ for the unitary dynamics and $J_Q^{\theta}=4B^2t^2+4\gamma t$ for the correlated parametrization. Here $B=0.1$, $\gamma=0.05$, $t=5$.}
  \label{fig.simultheta}
\end{figure}

In Fig.\ref{fig.simultheta}, we simulate the update of $\hat{\theta}$. The mean square error is simulated with $1000$ data obtained from repeating the adaptive procedure $1000$ times. For comparison, we also simulate the estimation under the unitary dynamics with the optimal adaptive control(the procedure is the same except replacing the adaptive QEC with the optimal adaptive control, which is given in Sec I of the supplemental material). It can be seen that precision of the correlated parametrization outperforms the unitary parametrization. The highest precision determined by the quantum Cramer-Rao bound, $\delta\hat{\theta}^2\geq \frac{1}{mKJ_Q^{\theta}}$, is also marked as the benchmark, which is obtained with the total number of measurements in the adaptive procedure ($mK$) and the maximal $J_Q^{\theta}$ achievable in the asymptotic limit, i.e., $J_Q^{\theta}=4B^2t^2$ for the unitary dynamics and $J_Q^{\theta}=4B^2t^2+4\gamma t$ for the correlated parametrization. 

\begin{figure}
 \includegraphics[width=0.55\textwidth]{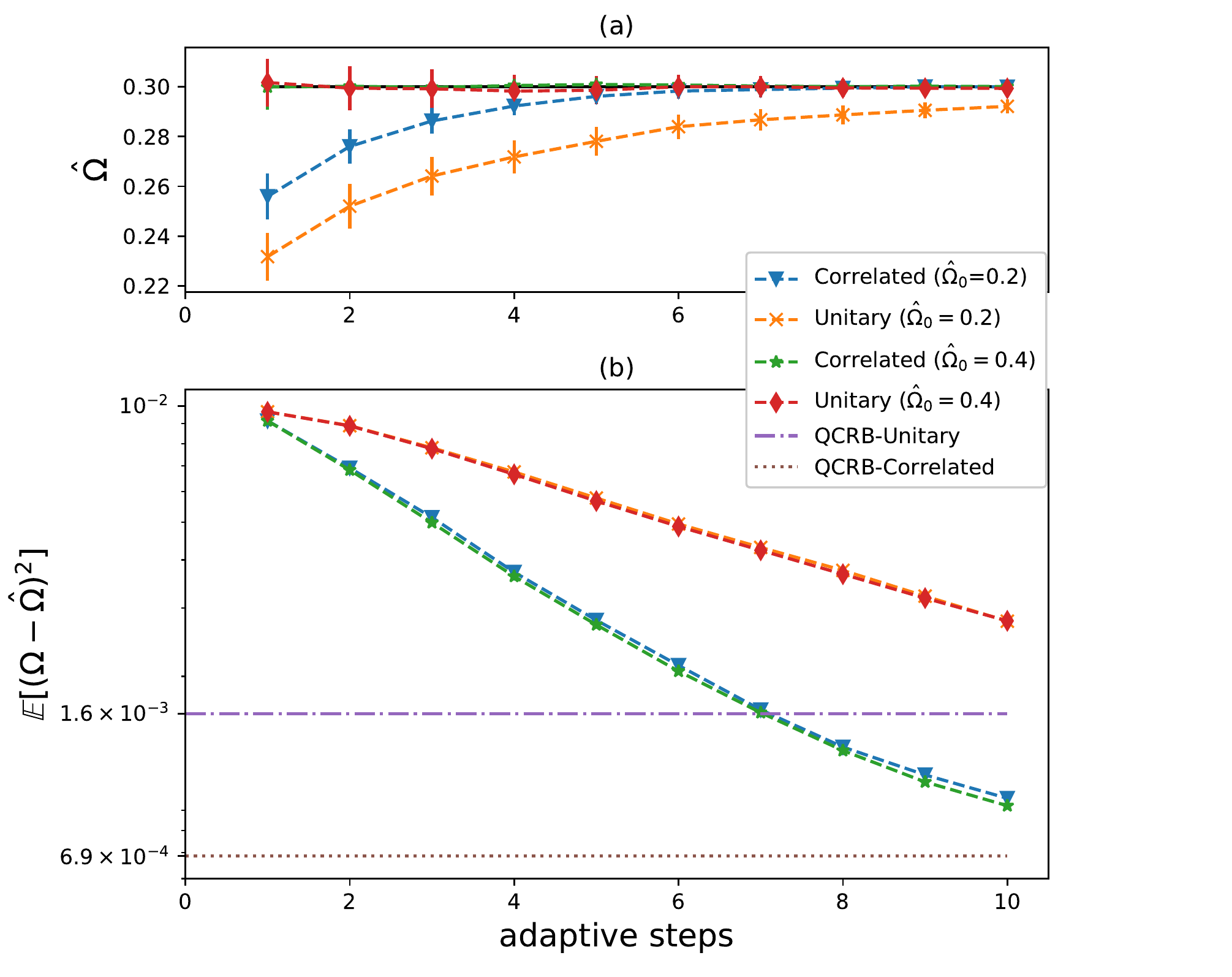}
 \caption{\textbf{Simulation for the estimation of $\Omega$.} (a) Simulation of the update of the estimator with the initial guess as $0.2$ and $0.4$ respectively, where the true value is taken as $\Omega=0.3$. The estimator under the optimally controlled unitary dynamics is also simulated for comparison. (b)Simulation of the mean square error of the maximal likelihood estimator under the correlated parametrization with the adaptive QEC and the unitary evolution with the optimal adaptive control respectively. The highest precision determined by the QCRB, $\delta \hat{\Omega}^2\geq \frac{1}{mK J_Q^{\Omega}}$, is also marked as the benchmark, where $mK=100$ is the total number of the measurements, $J_Q^{\Omega}$ is taken as the maximal QFI when $d\Omega=0$, i.e., $J_Q^{\Omega}=B^2t^4$ under the unitary dynamics and $J_Q^{\Omega}=B^2t^4+\frac{4\gamma t^3}{3}$ under the correlated parametrization. Here $B=0.1$, $\gamma=0.05$, $t=5$.}
 \label{fig:omega}
\end{figure}

The simulation for the estimation of $\Omega$ is similarly performed and shown in Fig.\ref{fig:omega} where the precision under the unitary evolution with the optimal adaptive control\cite{Pang2017} is also shown for comparison. It can be seen that the correlated parametrization outperforms the unitary evolution. The code of the simulation can be found on Github\footnote{https://github.com/anschen1994/CorEnhance}.

\section{General correlated parametrization}\label{sec:general}
We now consider the general correlated parametrization where the dynamics can be described by the master equation
\begin{equation}
\aligned
    \frac{d\rho}{dt} = & -i[H(x), \rho] \\ &+ \sum_{k=1}^m [E_k(x)\rho E_k^{\dag}(x) - \frac{1}{2}\{E_k^\dag(x)E_k(x), \rho\}],
\endaligned
\end{equation}
here $\{E_k(x)|k=1,\cdots, m\}$ are the Lindblad operators, $x$ is the parameter to be estimated. The adaptive QEC will be added as the control strategy. Again different from previous studies where the purpose of the QEC is to eliminate the noises and restore the noisy dynamics to the unitary evolution completely (the precision is thus at best the same as those under the unitary evolution)\cite{Plenio2000,Dur2014,Arrad2014,Kessler2014,Ozeri2013,Unden2016,Sekatski2017,Rafal2017,Zhou2018}, with the correlated parametrization the aim of the adaptive QEC is to eliminate the common part of the noises but keep the different part, this keeps the positive contributions of the noises from being smeared by their negative effects. We now show the general protocol.

Denote the space of a QEC code as $C$, which typically depends on $x$ and satisfies the QEC condition
\begin{align}
 \Pi_C(x) E_k(x) \Pi_C(x)&= \alpha_k\Pi_C(x), \\
 \Pi_C(x) E_k^{\dag}(x)E_j(x)\Pi_C(x) &= \beta_{kj}\Pi_C(x),
\end{align}
here $\Pi_C(x)$ denotes the projection on the code space\cite{Rafal2017,Zhou2018}. For the conventional schemes where the noises are independent of the parameter, the code space can also be independent of the parameter. With the correlated parametrization, however, the code space typically depends on the parameter. Since we do not know the value of the parameter a-priory, we can only design the code according to the estimated value, $\hat{x}$, obtained from the accumulated measurement data and update it adaptively. In the asympotic limit the estimation converges to the true value, thus
 \begin{align}\label{eq:QECcond}
  \lim_{\hat{x} \to x}\Pi_C(\hat{x}) E_k(x) \Pi_C(\hat{x}) &= \alpha_k\Pi_C(\hat{x}), \\
 \lim_{\hat{x} \to x}\Pi_C(\hat{x})E_k^{\dag}(x)E_j(x)\Pi_C(\hat{x}) &= \beta_{kj}(t)\Pi_C(\hat{x}).
\end{align}
We now derive the precision limit achievable in the asymptotic limit, i.e.,  when $\hat{x}\rightarrow x$. Suppose at time $t$, the probe state is in the code space, $\rho_C(t) = \Pi_C(\hat{x})\rho_C(t)\Pi_C(\hat{x})$(from now on we will also use $\Pi_C$ as a short notation for $\Pi_C(\hat{x})$ unless it is necessary to distinct $\hat{x}$ from $x$), after a small $dt$, the states evolves to
\begin{equation}
\aligned
    \rho(t+dt)& = \rho_C(t) - i[H(x), \rho_C(t)]dt \\
    & + \sum_{k=1}^m[ E_k(x)\rho_C(t)E_k^{\dag}(x) \\ & - \frac{1}{2}\{E_k^{\dag}(x)E_k(x),\rho_C(t)\}]dt
\endaligned
\end{equation}
A projection with $\Pi_C$ and $\Pi_E = I -\Pi_C$ is then performed on the state, which gives
\begin{equation}
\aligned
 &\lim_{\hat{x}\to x}\Pi_C(\hat{x})\rho(t+dt)\Pi_C(\hat{x})\\ &=\rho_C(t) -i[\Pi_CH(x)\Pi_C,\rho_C(t)] +\sum_{k=1}^m(|\alpha_k|^2-\beta_{kk})\rho_C(t),   \endaligned
 \end{equation}
\begin{equation}
\aligned
 &\lim_{\hat{x}\to x}\Pi_E(\hat{x})\rho(t+dt)\Pi_E(\hat{x})\\ &= \sum_k \Pi_E E_k(x)\rho_C(t)E_k^{\dag}(x)\Pi_E dt\\
 &=\sum_{k=1}^m [I-\Pi_C]E_k(x)\Pi_C\rho_C(t) \Pi_CE_k^{\dag}(x)[I-\Pi_C]dt\\
&=\sum_{k=1}^m M_k(x)\rho_C(t)M_k^\dagger(x),
\endaligned
\end{equation}
where we have used the facts that $\rho_C(t)\Pi_E=\Pi_E\rho_C(t) = 0$ for any state in the code space, here $M_k(x)=[I-\Pi_C]E_k(x)\Pi_C\sqrt{dt}$ describes the noisy effects that drive the state out of the code space.
Since
\begin{eqnarray}
\aligned
&\lim_{\hat{x}\rightarrow x}M_k^\dagger(x) M_k(x)\\
=&\lim_{\hat{x}\rightarrow x}\Pi_CE_k^\dagger(x)(I-\Pi_C)E_k(x)\Pi_C dt\\
=&\lim_{\hat{x}\rightarrow x}[\Pi_C E_k^\dagger(x) E_k(x)\Pi_C-\Pi_CE_k^\dagger(x)\Pi_C E_k(x)\Pi_C]dt\\
=&\lim_{\hat{x}\rightarrow x}[\beta_{kk}-|\alpha_k|^2]\Pi_C(\hat{x})dt\\
=&\lim_{\hat{x}\rightarrow x} d_{kk}\Pi_C(\hat{x})dt,
\endaligned
\end{eqnarray}
here $d_{kk}=\beta_{kk}-|\alpha_k|^2$. The polar decomposition of $M_k(x)$ can thus be written as $M_k(x)=\sqrt{d_{kk}}U_k(x)\Pi_C(\hat{x})\sqrt{dt}$.
Without loss of generality, we can assume the errors are orthogonal, i.e., $M_k^\dagger M_j=\delta_{kj}d_{kk}\Pi_Cdt$(nonorthogonal noise can be made orthogonal by equivalent unitary transformation as in the standard quantum error correction). Since $\Pi_C(I-\Pi_C)=\Pi_C-\Pi_C=0$, we also have $\Pi_C M_k=0$, thus $\Pi_CU_k\Pi_C=0$.

The state after the projection of $\{\Pi_C,\Pi_E\}$ can now be written as
\begin{equation}
\aligned
\rho_p(t+dt)=&\Pi_C\rho(t+dt)\Pi_C+\sum_{k=1}^m M_k\rho(t+dt) M_k^\dagger\\
=&\Pi_C\rho(t+dt)\Pi_C\\
&+\sum_{k=1}^m d_{kk}U_k\Pi_C\rho(t+dt)\Pi_CU_k^{\dagger}dt
\endaligned
\end{equation}
The recovery operation, $R_E$, can be taken as the operation with the Kraus operators consisting of  $\{\Pi_C,\Pi_CU_1^{\dagger}(\hat{x}),\cdots,\Pi_CU_m^{\dagger}(\hat{x})\}$, where $\hat{U}_k^{\dag}(\hat{x})$ is obtained from the estimated error operator $M_k(\hat{x})=\Pi_E(\hat{x})E_k(\hat{x})\Pi_C(\hat{x})\sqrt{dt}=\sqrt{d_{kk}}\hat{U}_k(\hat{x})\Pi_C(\hat{x})\sqrt{dt}$. The recovered state at $t+dt$ is then given by $\rho_C(t+dt)=R_E[\rho_p(t+dt)]$. From the difference between $\rho_C(t+dt)$ and $\rho_C(t)$, we can then obtain the dynamics for the recovered state, $\frac{d \rho_C}{dt}=L_x^C(\rho_C)$. In the supplemental material we show that the dynamics $L_x^C$ can be expanded around $\hat{x}$ as $L_x^C=L_0+L_1dx+L_2dx^2+O(dx^3)$, with $dx=x-\hat{x}$, $L_0=L_{\hat{x}}^C$, $L_1=\partial_x L_{x}^C|_{x=\hat{x}}$, and $L_2=\frac{1}{2}\frac{\partial^2L_{x}^C}{\partial x^2}|_{x=\hat{x}}$. Specifically,
\begin{eqnarray}
\label{eq:expansionO}
\aligned
&L_0(\rho)=-i[\Pi_C(\hat{x}) H(\hat{x}) \Pi_C(\hat{x}),\rho],\\
&L_1(\rho)=-i[\Pi_C(\dot{H}+\frac{i}{2}\sum_kE_k^\dagger\dot{E}_k-\dot{E}_k^\dagger E_k)\Pi_C, \rho],\\
&L_2(\rho)=-i\frac{1}{2}[\Pi_C(\ddot{H}+\frac{i}{2}\sum_k E_k^{\dag}\ddot{E}_k-\ddot{E}_k^{\dag}E_k)\Pi_C,\rho] \\
&+\sum_k[\Pi_C\dot{E}_{k}\Pi_C\rho\Pi_C\dot{E}_{k}^\dagger\Pi_C-\frac{1}{2}\{\Pi_C\dot{E}_{k}^{\dag}\dot{E}_k\Pi_C,\rho\}]\\
&+\sum_{k,j}\frac{1}{d_{kk}}\Pi_C(E_k^\dagger\dot{E}_j-\alpha_k^*\dot{E}_j)\Pi_C\rho\Pi_C(\dot{E}_j^\dagger E_k-\alpha_k\dot{E}_j^\dagger)\Pi_C
\endaligned
\end{eqnarray}
where we use the over-dot to denote the derivatives with respect to $x$ which is then evaluated at the point $x=\hat{x}$, for example, $\ddot{H}=\frac{\partial^2 H}{\partial x^2}|_{x=\hat{x}}$. We note that $L_0(\rho)=-i[\Pi_C(\hat{x}) H(\hat{x}) \Pi_C(\hat{x}),\rho]$ is a unitary evolution, where $\Pi_C(\hat{x})H(\hat{x})\Pi_C(\hat{x})$ only depends on $\hat{x}$ and does not contribute to the precision. It can be cancelled by an additional control Hamiltonian taken as $H_C=-\Pi_C(\hat{x})H(\hat{x})\Pi_C(\hat{x})$. Thus without loss of generality, we can take $L_0$ as zero. The dynamics, up to the second order of $dx$, is then
\begin{equation}
\label{eq:QECdynamics}
\frac{d\rho_C}{dt}=[L_1dx+L_2dx^2](\rho_C).
\end{equation}

The second order expansion is sufficient for the computation of QFI, as from Eq.(\ref{eq:Bures}) we know the QFI is determined by the distance between two neighboring states, $\rho_C(x)$ and $\rho_C(x+dx)$, up to the second of $dx$. In the asymptotic limit when $\hat{x}=x$, $\rho_C(x)$ can be obtained from equation (\ref{eq:QECdynamics}) with $dx=0$, which gives $\rho_C(x)=\rho_0$. While $\rho_C(x+dx)$, up to the second order of $dx$, can be obtained as
\begin{eqnarray}
\aligned
&\rho_C(x+dx)\\
=&e^{(L_1dx+L_2dx^2)t}(\rho_0)\\
=&\left[I+tL_1 dx+tL_2dx^2+\frac{1}{2}t^2L_1^2dx^2+O(dx^3)\right](\rho_0).
\endaligned
\end{eqnarray}
We note that the terms involving the commutators between $L_1$ and $L_2$ contain higher orders($\geq 3$) of $dx$ thus does not contribute to the QFI.
If we take the initial probe state as $\rho_0=|\psi\rangle\langle\psi|$, then
the fidelity between $\rho_C(x)$ and $\rho_C(x+dx)$ can be obtained as
\begin{eqnarray}
\nonumber
\aligned
&F^2(\rho_{x},\rho_{x+dx})=\langle\psi|\rho_C(x+dx)|\psi\rangle\\
=&1+tdx \langle\psi|L_1(|\psi\rangle\langle\psi|)|\psi\rangle+\frac{1}{2}t^2dx^2\langle\psi|L_1^2(|\psi\rangle\langle\psi|)|\psi\rangle\\
&+tdx^2\langle\psi|L_2(|\psi\rangle\langle\psi|)|\psi\rangle+O(dx^3).
\endaligned
\end{eqnarray}
Using equation (\ref{eq:expansionO}) it is easy to obtain
\begin{eqnarray}
\aligned
\langle\psi|L_1(|\psi\rangle\langle\psi|)|\psi\rangle=&0,\\
\langle\psi|L_1^2(|\psi\rangle\langle\psi|)|\psi\rangle=&-2[\langle\psi|\widetilde{H}^2|\psi\rangle-\langle\psi|\widetilde{H}|\psi\rangle^2],\\
\endaligned
\end{eqnarray}
where $\widetilde{H}=\Pi_C[\dot{H}+\frac{i}{2}\sum_k(E_k^\dagger\dot{E}_k-\dot{E}_k^\dagger E_k)]\Pi_C$. 
From equation (\ref{eq:Bures}) we can then obtain the QFI as
\begin{equation}
\label{eq:QFIQEC}
\aligned
J_Q^x&=\lim_{dx\rightarrow 0}\frac{8-8F[\rho_C(x),\rho_C(x+dx)]}{dx^2}\\
&=4t^2\Delta_{|\psi\rangle}^2(\widetilde{H})-4t\langle\psi|L_2(|\psi\rangle\langle\psi|)|\psi\rangle,
\endaligned
\end{equation}
where $\Delta_{|\psi\rangle}^2(\widetilde{H})=\langle\psi|\widetilde{H}^2|\psi\rangle-\langle\psi|\widetilde{H}|\psi\rangle^2$. The second term here is always non-negative, i.e., $-\langle\psi|L_2(|\psi\rangle\langle\psi|)|\psi\rangle\geq 0$. This can be seen by looking at the changes of $Tr(\rho_C\rho_0)$ with $\rho_0=|\psi\rangle\langle\psi|$. As $\frac{d \rho_C}{dt}=L_x^C(\rho_C)$, we have
\begin{eqnarray}
\aligned
    \frac{d Tr(\rho_C\rho_0)}{dt}&=Tr(\frac{d \rho_C}{dt}\rho_0)\\
    &=\langle\psi|\frac{d \rho_C}{dt}|\psi\rangle\\
    &=\langle\psi|L_x^C(\rho_C)|\psi\rangle\\
    &=\langle\psi|L_0(\rho_C)|\psi\rangle+\langle\psi|L_1(\rho_C)|\psi\rangle dx\\
    &+\langle\psi|L_2(\rho_C)|\psi\rangle dx^2+O(dx^3)
\endaligned
\end{eqnarray}
At $t=0$, $\rho_C=\rho_0$, $Tr(\rho_C\rho_0)=1$, which is the maximal value $Tr(\rho_C\rho_0)$ can take and it can only decrease, i.e., $\frac{d Tr(\rho_C\rho_0)}{dt}\leq 0$ at $t=0$. Note that both $L_0(\rho_C)$ and $L_1(\rho_C)$ are in the form of $-i[A,\rho_C]$, thus when $\rho_C=\rho_0=|\psi\rangle\langle\psi|$ at $t=0$, we have
\begin{eqnarray}
\langle\psi|-i[A,|\psi\rangle\langle\psi|]|\psi\rangle=-i[\langle\psi|A|\psi\rangle-\langle\psi|A|\psi\rangle]=0.
\end{eqnarray}
At $t=0$, $\langle\psi|L_2(|\psi\rangle\langle\psi|)|\psi\rangle$ is then the leading term whose sign should be the same with $\frac{d Tr(\rho_C\rho_0)}{dt}\leq 0$. This shows $-\langle\psi|L_2(|\psi\rangle\langle\psi|)|\psi\rangle\geq 0$.

We now connect this general formula to the result in Sec.\ref{sec:direction}. For the example of estimating $\theta$, the dynamics is given by
\begin{equation}
\frac{d \rho}{d t} = -i[B\sigma_{\vec{n}(\theta)},\rho]+\gamma [\sigma_{\vec{n}(\theta)}\rho \sigma_{\vec{n}(\theta)} - \rho]
\end{equation}
with $\sigma_{\vec{n}(\theta)}=\cos\theta\sigma_1+\sin\theta\sigma_3$. In this case $H(\theta)=B\sigma_{\vec{n}(\theta)}$ and we have one Lindblad operator, $E(\theta)=\sqrt{\gamma}\sigma_{\vec{n}(\theta)}$. With an ancilla spin the code space is chosen as $\{|C_0\rangle=|+_{\hat{\theta}}\rangle|+_{\hat{\theta}}\rangle, |C_1\rangle=|-_{\hat{\theta}}\rangle|-_{\hat{\theta}}\rangle\}$, where $|+_{\hat{\theta}}\rangle=-\cos\frac{\hat{\theta}}{2}|0\rangle + \sin\frac{\hat{\theta}}{2}|1\rangle$, $|-_{\hat{\theta}}\rangle=\sin\frac{\hat{\theta}}{2}|0\rangle + \cos\frac{\hat{\theta}}{2}|1\rangle$ are the eigenvectors of $\frac{\partial H(\hat{\theta})}{\partial \theta}=B(-\sin \hat{\theta}\sigma_1+\cos \hat{\theta}\sigma_3)$, together   with $|C_2\rangle=\sigma_{\vec{n}(\hat{\theta})}|C_0\rangle=|-_{\hat{\theta}}\rangle|+_{\hat{\theta}}\rangle$ and $|C_3\rangle=\sigma_{\vec{n}(\hat{\theta})}|C_1\rangle=|+_{\hat{\theta}}\rangle|-_{\hat{\theta}}\rangle$, they form a basis for two spins. We then have $\Pi_C=|C_0\rangle\langle C_0|+|C_1\rangle\langle C_1|$.

To obtain $L_0$, note that
\begin{eqnarray}
\aligned
&\Pi_CH(\hat{\theta})\Pi_C\\
&=(|C_0\rangle\langle C_0|+|C_1\rangle\langle C_1|)B\sigma_{\vec{n}(\hat{\theta})}(|C_0\rangle\langle C_0|+|C_1\rangle\langle C_1|)\\
&=B(|C_0\rangle\langle C_0|+|C_1\rangle\langle C_1|)(|C_2\rangle\langle C_0|+|C_3\rangle\langle C_1|)\\
&=\textbf{0}.
\endaligned
\end{eqnarray}
Thus $L_0=\textbf{0}$, there is no need to add an additional control to cancel it.

To obtain $L_1$, note that $\dot {H}+\frac{i}{2}(E^\dagger\dot{E}-\dot{E}^\dagger E)=\frac{\partial H(\hat{\theta})}{\partial \theta}+\gamma \sigma_2$,then
\begin{eqnarray}
 \aligned
 &\widetilde{H}=\Pi_C [\dot {H}+\frac{i}{2}(E^\dagger\dot{E}-\dot{E}^\dagger E)]\Pi_C\\
=& (|C_0\rangle\langle C_0|+|C_1\rangle\langle C_1|)\frac{\partial H(\hat{\theta})}{\partial \theta}(|C_0\rangle\langle C_0|+|C_1\rangle\langle C_1|)\\
&+(|C_0\rangle\langle C_0|+|C_1\rangle\langle C_1|)\gamma \sigma_2(|C_0\rangle\langle C_0|+|C_1\rangle\langle C_1|)\\
=&B(|C_0\rangle\langle C_0|+|C_1\rangle\langle C_1|)(|C_0\rangle\langle C_0|-|C_1\rangle\langle C_1|)\\
=&B(|C_0\rangle\langle C_0|-|C_1\rangle\langle C_1|),
 \endaligned
\end{eqnarray}
where we have used the facts that $\langle C_0|\sigma_2|C_1\rangle=0$, $\frac{\partial H(\hat{\theta})}{\partial \theta}|C_0\rangle=B|C_0\rangle$, $\frac{\partial H(\hat{\theta})}{\partial \theta}|C_1\rangle=-B|C_1\rangle$.
If we prepare the initial probe state as $|\psi\rangle =\frac{|C_0\rangle+|C_1\rangle}{\sqrt{2}}$, then
\begin{eqnarray}
\aligned
\Delta_{|\psi\rangle}^2(\widetilde{H})&=\langle\psi|\widetilde{H}^2|\psi\rangle-\langle\psi|\widetilde{H}|\psi\rangle^2\\
&=B^2.
\endaligned
\end{eqnarray}

To obtain $L_2$, note that $\ddot{H}=\frac{\partial ^2H(\hat{\theta})}{\partial \theta^2}=-B\sigma_{\vec{n}(\hat{\theta})}$, $E=\frac{\sqrt{\gamma}}{B}H$, $\dot{E}=\frac{\sqrt{\gamma}}{B}\dot{H}$, $\ddot{E}=\frac{\sqrt{\gamma}}{B}\ddot{H}$. 
Substitute these into Eq.(\ref{eq:expansionO}) we obtain
\begin{equation}
    L_2(\rho)=\gamma(\sigma_3^C\rho\sigma_3^C-\rho),
\end{equation}
here $\sigma_3^C=|C_0\rangle\langle C_0|-|C_1\rangle\langle C_1|$. With the initial state $|\psi\rangle=\frac{|C_0\rangle+|C_1\rangle}{\sqrt{2}}$, 
$\langle\psi|L_2(|\psi\rangle\langle\psi|)|\psi\rangle=-\gamma.$

The QFI that can be achieved in the asymptotic limit when $\hat{\theta}\rightarrow \theta$ can then be obtained from Eq.(\ref{eq:QFIQEC}) directly, which is
\begin{equation}
\aligned
J_Q^{\theta}=4B^2t^2 + 4\gamma t.
\endaligned
\end{equation}
This is the same as we obtained previously.

\section{Conclusion}\label{sec:discussion}
\noindent It is widely believed that the precision at the presence of the noise is always worse than what can be achieved under the unitary dynamics. Our study, however, provides a new perspective on the role of noises in quantum metrology. By showing that the fluctuation, one of the main noises in quantum dynamics, can actually help improving the precision for some tasks in quantum metrology, we changed the shared belief. This also changes our understanding on the ultimate precisions achievable in quantum metrology. Our study suggests that, instead of trying to suppress the noises, for correlated parametrization it can actually be beneficial to increase the noise---sometimes the stronger the noise, the higher the precision. This differs from previous studies in quantum computation that exploit noises, such as environmental engineering and noise-assisted quantum error correction, where the environment(or the noise) needs to be explicitly controlled or designed and typically can not outperform the unitary dynamics. We expect our study will lead to the explorations of the correlated parametrization in many applications and discover new ultimate precision limits beyond what were believed to be achievable previously. Future studies include the optimization of the control strategies, extension to multi-parameter quantum estimation and purposely engineer the correlated parametrization for various applications.

\appendix

\section{Precision limits under the unitary dynamics with the optimal adaptive control}
We provide the highest precisions achievable under the unitary dynamics with the optimal adaptive control, which are used for comparison.

For the estimate of $\theta$ under the unitary dynamics
\begin{equation}
\frac{d|\psi\rangle}{dt}=-iB\sigma_{\vec{n}(\theta)}|\psi\rangle,
\end{equation}
where $\sigma_{\vec{n}(\theta)}=\cos\theta\sigma_1+\sin\theta\sigma_3$, it is known that the optimal adaptive control is to reverse the dynamics, which can be achieved by adding a control Hamiltonian $H_c=-B\sigma_{\vec{n}(\hat{\theta})}$ with $\hat{\theta}$ as the estimated value\cite{yuan2015optimal,Braun2017}. In the asymptotical limit when $\hat{\theta}$ converges to $\theta$, it achieves the highest QFI as $4B^2t^2$\cite{yuan2015optimal}. In the finite regime where $\hat{\theta}$ may not equal to $\theta$, the effective Hamiltonian is given by
\begin{equation}
\label{eq:H}
H_{\mathrm{eff}}=B(\sigma_{\vec{n}}-\hat{\sigma}_{\vec{n}})=B[(\cos\theta-\cos\hat{\theta})\sigma_1+(\sin\theta-\sin\hat{\theta})\sigma_3],
 \end{equation}
the dynamics is given by $U_{\mathrm{eff}}(t) = e^{-iB(\sigma_{\vec{n}}-\hat{\sigma}_{\vec{n}})t}$. With the optimal probe state, which is the maximal entangled state, $|\psi(0)\rangle = \frac{|00\rangle + |11\rangle}{\sqrt{2}}$, the QFI is
\begin{eqnarray}
\aligned
J_Q^{\theta}= \frac{1}{2}[&1+4B^2t^2+4B^2t^2\cos(\theta-\hat{\theta})\\
&-\cos(2Bt\sqrt{2-2\cos(\theta-\hat{\theta})})].
\endaligned
\end{eqnarray}
This can be expanded up to the fourth order of $d\theta=\hat{\theta}-\theta$ as
\begin{equation}
J_Q^{\theta} = 4B^2t^2 - \frac{1}{3}B^4t^4d\theta^4 + o(d\theta^4).
\end{equation}
The $t^2$ scaling is maintained within the time order of $(Bd\theta)^{-1}$. We note that this is not much a restriction, in anyway the evolution time should be restricted within this time order in order to avoid the phase ambiguity. Intuitively, up to the order of $d\theta$, the effective Hamiltonian in Eq.(\ref{eq:H}) is approximately $H_{\mathrm{eff}}\approx Bd\theta\sigma_3$, the evolution time should be restricted to $Bd\theta t\leq 2\pi$ in order to avoid the phase ambiguities(phases that differ $2\pi$ can not be differentiated), which restricts the evolution time to the order of $(Bd\theta)^{-1}$. Beyond this time order, it may not be possible to obtain a unique estimation.

For the estimation of $\Omega$ under the unitary dynamics,
\begin{equation}
\frac{d|\psi\rangle}{dt}=-iB\sigma_{\vec{n}}(\Omega, t)|\psi\rangle,
\end{equation}
with $\sigma_{\vec{n}}(\Omega,t)=\cos(\Omega t) \sigma_1-\sin(\Omega t) \sigma_2$, the optimal adaptive control Hamiltonian can be taken as $H_c=-B\sigma_{\vec{n}}(\hat{\Omega},t)+\frac{\hat{\Omega}}{2}\sigma_3$ with $\hat{\Omega}$ as the estimated value\cite{Pang2017}. In the asymptotical limit when $\hat{\Omega}$ converges to $\Omega$, it achieves the maximal QFI, $J_Q^{\Omega}=B^2t^4$\cite{Pang2017}. In the finite regime when $\hat{\Omega}$ may not equal to $\Omega$, the QFI is approximately
\begin{equation}
J_Q^{\Omega} = B^2t^4(1-\frac{1}{18}t^2d\Omega^2)
\end{equation}
up to the second order of $d\Omega=\hat{\Omega}-\Omega$\cite{Pang2017}.

Here the asymptotical limit refers to the repetition of the experiment. When the experiment is repeated with sufficient times the estimation converges to the true value(if the experiment is repeated with $n$ times, the QFI is $nJ_Q$). For each experiment, the evolution time $t$ is a finite value and $4B^2t^2$ ($B^2t^4$) is believed to the highest QFI that can be achieved within the given $t$ for the estimation of $\theta$($\Omega$).

In practice, the controls are applied adaptively. The experiment is first repeated for a number of times, an estimation is then made based on the collected data. The controls are then updated according to the estimated value then new data are collected. This is repeated for a number of rounds with the QFI approaching the highest value when the estimation converges to the true value.

\section{Estimation with the average of the fluctuation}
We derive the precision at the presence of the fluctuation under the free evolution. We will obtain the precision with two different approaches, one from the quantum trajectory and the other from the equivalent master equation.

In the main text we showed that for the estimation of $B$ under the fluctuating field,
\begin{equation}
\frac{d|\psi\rangle}{dt}=-i[B+\xi(t)]\sigma_{\vec{n}}|\psi\rangle,
\end{equation}
along each trajetory $U(t)=e^{-i[Bt+\int_0^t \xi(\tau)d\tau]\sigma_{\vec{n}(\theta)}}$,
\begin{eqnarray}
\aligned
h_B&=\int_0^tU^\dagger(\tau)\sigma_{\vec{n}(\theta)}U(\tau)d\tau\\
&=t\sigma_{\vec{n}(\theta)}.
\endaligned
\end{eqnarray}
The QFI is thus upper bounded by $4t^2$ along each trajectory. However, as the QFI is a statistical quantity which is only meaningful over many repetitions. We give an analysis on the precision limit under the repetitions.

For each trajectory, the optimal probe state can be taken as $|\psi(0)\rangle=\frac{|\lambda_{\max}\rangle+|\lambda_{\min}\rangle}{\sqrt{2}}$, where $|\lambda_{\max/\min}\rangle$ is the eigenvector of $\sigma_{\vec{n}(\theta)}$(for the estimation of $B$, $\theta$ is assumed to be known, the eigenvectors are thus also known, and if an ancilla spin is available the optimal probe state can also be taken as $|\psi(0)\rangle=\frac{|00\rangle+|11\rangle}{\sqrt{2}}$), and the optimal measurement can be taken as the projective measurements on the two states, $\frac{|\lambda_{\max}\rangle\pm e^{2i\beta}|\lambda_{\min}\rangle}{\sqrt{2}}$ (or $\frac{|\lambda_{\max}\rangle|\lambda_{\max}\rangle\pm e^{2i\beta}|\lambda_{\min}\rangle|\lambda_{\min}\rangle}{\sqrt{2}}$ if the probe state is $\frac{|00\rangle+|11\rangle}{\sqrt{2}}$). The probability for the measurement outcomes are $p_1=\cos^2[Bt+\int_0^t \xi(\tau)d\tau+\beta]$ and $p_2=\sin^2[Bt+\int_0^t \xi(\tau)d\tau+\beta]$. For a fixed trajectory, this has the classical Fisher information, $J_C^B=\frac{1}{p_1}(\frac{\partial p_1}{\partial B})+\frac{1}{p_2}(\frac{\partial p_2}{\partial B})=4t^2$. However, in practice, the probability of the measurement outcomes can only be obtained by repeating the procedure many times, and each time the realization of the fluctuation is different. The actually probability is thus the average over many repetitions, i.e., $p_1=\int_{-\infty}^{+\infty} \cos^2[Bt+\phi+\beta]p(\phi)d\phi$, here $\phi=\int_0^t \xi(\tau)d\tau$ which has a Gaussian distribution with $E(\phi)=0$ and $E(\phi^2)=\gamma t$. Thus over many realizations we have
\begin{eqnarray}
\aligned
p_1=& \int_{-\infty}^{+\infty} \cos^2[Bt+\phi+\beta]\frac{1}{\sqrt{2\pi \gamma t}}e^{-\frac{\phi^2}{2\gamma t}}d\phi\\
=&\int_{-\infty}^{+\infty} \frac{1+\cos(2Bt+2\phi+2\beta)}{2}\frac{1}{\sqrt{2\pi \gamma t}}e^{-\frac{\phi^2}{2\gamma t}} d\phi\\
=&\frac{1+e^{-2\gamma t}\cos(2Bt+2\beta)}{2},
\endaligned
\end{eqnarray}
where we made use of the formulas $\int_{-\infty}^{+\infty} e^{-a\phi^2}\cos(k\phi)d\phi=\sqrt{\frac{\pi}{a}}e^{\frac{-k^2}{4a}}$ and $\int_{-\infty}^{+\infty} e^{-a\phi^2}\sin(k\phi)d\phi=0$. $p_2=1-p_1=\frac{1-e^{-2\gamma t}\cos(2Bt+2\beta)}{2}$. The classical Fisher information achieves the maximal value $J_C^B=4t^2e^{-4\gamma t}$ when $\beta=\frac{\pi}{4}-Bt$, here $\beta$ actually depends on $B$, which means that the optimal measurement can only be realized adaptively. In practice, $\beta$ is taken as $\frac{\pi}{4}-\hat{B}t$ with $\hat{B}$ as the estimaed value obtained from previous data, the measurement then converges to the optimal measurement in the asymptotical limit when $\hat{B}$ converges to $B$.

It is easy to check that under the equivalent master equation
\begin{equation}
\frac{d \rho}{d t} = -i[B\sigma_{\vec{n}(\theta)},\rho]+\gamma [\sigma_{\vec{n}(\theta)}\rho \sigma_{\vec{n}(\theta)} - \rho],
\end{equation}
with the optimal probe state $|\psi(0)\rangle=\frac{|\lambda_{\max}\rangle+|\lambda_{\min}\rangle}{\sqrt{2}}$ or $|\psi(0)\rangle=\frac{|00\rangle+|11\rangle}{\sqrt{2}}$, the QFI is exactly $4t^2e^{-4\gamma t}$, which is consistent with the above analysis. 

For the estimation of $\theta$ under the free evolution, with an ancillary spin we can prepare the probe state as $|\psi(0)\r = \frac{|00\rangle + |11\rangle}{\sqrt{2}}$. Along one trajectory the state at time $t$ is then $|\psi(t)\rangle = (e^{-i\Phi\sigma_n}\otimes \I)\frac{|00\rangle +|11\rangle }{\sqrt{2}}$, where $\Phi = Bt + \int_0^t \xi(\tau)d\tau$. We then perform the projective measurement on the Bell basis$(\frac{|00\rangle+|11\rangle}{\sqrt{2}},\frac{|00\rangle-|11\rangle}{\sqrt{2}},\frac{|10\rangle+|01\rangle}{\sqrt{2}},\frac{|10\rangle-|01\rangle}{\sqrt{2}})$, which has the probability distribution as
\begin{equation}
\aligned
& p_1 = \cos^2\Phi, \\
& p_2 = \sin^2\Phi \sin^2\theta, \\
& p_3 = \sin^2\Phi \cos^2\theta, \\
& p_4 = 0.
\endaligned
\end{equation}
With many repetitions the actual probabilities are given by $\bar{p}_i = \int_{-\infty}^{+\infty}p_id\Phi$, where $\Phi$ has a Gaussian distribution with $E[\Phi]=Bt$ and $E[(\Phi-Bt)^2]=\gamma t$. We then have
\begin{equation}
\aligned
& \bar{p}_1 = \frac{1}{2}\left(1+e^{-2\gamma t}\cos 2Bt\right), \\
& \bar{p}_2 = \frac{1}{2}\sin^2\theta(1-e^{-2\gamma t}\cos 2Bt), \\
& \bar{p}_3 = \frac{1}{2}\cos^2\theta(1-e^{-2\gamma t}\cos2Bt), \\
& \bar{p}_4 = 0.
\endaligned
\end{equation}
The classical Fisher information under the free evolution is then
\begin{equation}
\aligned
J_C^{\theta} = \sum_{i=1}^4 \frac{(\partial_{\theta}\bar{p}_i)^2}{\bar{p}_i} = 2(1-e^{-2\gamma t}\cos 2Bt).
\endaligned
\end{equation}
We can also derive QFI from the master equation
\begin{equation}
\frac{d \rho}{d t} = -i[B\sigma_{\vec{n}(\theta)},\rho]+\gamma [\sigma_{\vec{n}(\theta)}\rho \sigma_{\vec{n}(\theta)} - \rho]
\end{equation}
with the optimal probe state $|\psi(0)\r = \frac{|00\rangle + |11\rangle}{\sqrt{2}}$.
Under the free evolution the dynamics on the probe spin can be equivalent to a quantum channel with the Kraus operators $K_1(\theta)=\sqrt{\frac{1+\eta}{2}}e^{-i B\sigma_{\vec{n}(\theta)} t}$, $K_2(\theta)=\sqrt{\frac{1-\eta}{2}}\sigma_{\vec{n}(\theta)}e^{-i B\sigma_{\vec{n}(\theta)} t}$, where $\eta=e^{-2\gamma t}$.
Under free evolution the state at time $t$ is then $\rho(\theta)=K_1(\theta)\otimes I\left|\psi(0)\right>\left<\psi(0)\right|K_1^\dagger(\theta)\otimes I+K_2(\theta)\otimes I\left|\psi(0)\right>\left<\psi(0)\right|K_2^\dagger(\theta)\otimes I$ (here $I$ is the Identity operator that acts on the ancillary spin), which can be diagonalized as $\rho(\theta)=\sum_{i=1}^4\lambda_i\left|e_i\right>\left<e_i\right|$, with the eigenvalues $\lambda_1=\lambda_2=0$, $\lambda_3=\frac{1-\eta}{2}$, $\lambda_4=\frac{1+\eta}{2}$ and the corresponding eigenvectors
\begin{equation}
  \begin{aligned}
   & \left|e_1\right>=\frac{1}{\sqrt{2}}
  \begin{pmatrix}
    -\cos\theta \\ \sin\theta \\ \sin\theta \\ \cos\theta
  \end{pmatrix},
  \left|e_2\right>=\frac{1}{\sqrt{2}}
  \begin{pmatrix}
    0 \\ 1 \\ -1 \\ 0
  \end{pmatrix},\\
  &\left|e_3\right>=\frac{1}{\sqrt{2}}
  \begin{pmatrix}
    -i\sin(Bt)+\cos(Bt)\sin\theta \\
    \cos(Bt)\cos\theta \\
    \cos(Bt)\cos\theta \\
    -i\sin(Bt)-\cos(Bt)\sin\theta
  \end{pmatrix},\\
  &\left|e_4\right>=\frac{1}{\sqrt{2}}
  \begin{pmatrix}
    \cos(Bt)-i\sin(Bt)\sin\theta \\
    -i\sin(Bt)\cos\theta \\
    -i\sin(Bt)\cos\theta \\
    \cos(Bt)+i\sin(Bt)\sin\theta
  \end{pmatrix}.
  \end{aligned}
\end{equation}
From $\partial\rho(\theta)/\partial\theta=\frac{1}{2}\left(\rho(\theta)L_s(\theta)+L_s(\theta)\rho(\theta)\right)$, we can then obtain the SLD operator, $L_s(\theta)$ as $L(\theta)=-2\cos(Bt)(\left|e_1\right>\left<e_3\right|+\left|e_3\right>\left<e_1\right|)+2i\sin(Bt)(\left|e_1\right>\left<e_4\right|-\left|e_4\right>\left<e_1\right|)$.
The QFI can then be computed as $J_Q^\theta=\tr(\rho(\theta)L^2(\theta))=2(1-\eta\cos 2Bt)=2(1-e^{-2\gamma t}\cos 2Bt)$. This is consistent with the above analysis using the quantum trajectory. 

\section{Dynamics for general correlated parametrization under the adaptive QEC}
 In the main text we show that if the state at time $t$, denoted as $\rho_C(t)$, is in the code space, then at time $t+dt$ the state evolves to 
 \begin{equation}
\aligned
    \rho(t+dt)& = \rho_C(t) - i[H(x), \rho_C(t)]dt \\
    & + \sum_{k=1}^m E_k(x)\rho_C(t)E_k^{\dag}(x) \\ & - \frac{1}{2}\{E_k^{\dag}(x)E_k(x),\rho_C(t)\}dt.
\endaligned
\end{equation}
A projection consists of $\{\Pi_C,\Pi_E=I-\Pi_C\}$ is then performed to get 
\begin{equation}
\aligned
\rho_p(t+dt)=\Pi_C\rho(t+dt)\Pi_C+\Pi_E\rho(t+dt)\Pi_E,
\endaligned
\end{equation}
here \begin{equation}
\aligned
 &\Pi_E(\hat{x})\rho(t+dt)\Pi_E(\hat{x})
&=\sum_{k=1}^m M_k(x)\rho_C(t)M_k^\dagger(x)
\endaligned
\end{equation}
with $M_k(\hat{x})=\Pi_E(\hat{x})E_k(\hat{x})\Pi_C(\hat{x})\sqrt{dt}$. And we can write $M_k(\hat{x})=\sqrt{d_{kk}}\hat{U}_k(\hat{x})\Pi_C(\hat{x})\sqrt{dt}$ using the polar decomposition as shown in the main text. We then perform the recovery operation to get the corrected state at $t+dt$ as $\rho_C(t+dt)=R_E[\rho_p(t+dt)]$, where $R_E$ is the recovery operation with the Kraus operators consisting of  $\{\Pi_C,\Pi_CU_1(\hat{x}),\cdots,\Pi_CU_m(\hat{x})\}$, here $\hat{U}_k^{\dag}(\hat{x})$ is obtained from the estimated value of $x$.

Note that the recovery operation can also be applied directly on the state $\rho(t+dt)$, the projective measurement of $\{\Pi_C, \Pi_E\}$ can be skipped since
\begin{eqnarray}
\aligned
&R_E[\rho(t+dt)]\\
=&\Pi_C\rho(t+dt)\Pi_C+\sum_k\Pi_C\hat{U}_k^{\dag}(\hat{x})\rho(t+dt)\hat{U}_k(\hat{x})\Pi_C\\
=&\Pi_C\rho(t+dt)\Pi_C\\
+&\sum_k\Pi_C\hat{U}_k^{\dag}(\hat{x})(\Pi_C+\Pi_E)\rho(t+dt)(\Pi_C+\Pi_E)\hat{U}_k(\hat{x})\Pi_C\\
=&\Pi_C\rho(t+dt)\Pi_C+\sum_k\Pi_C\hat{U}_k^{\dag}(\hat{x})\Pi_E\rho(t+dt)\Pi_E\hat{U}_k(\hat{x})\Pi_C\\
=&R_E[\rho_p(t+dt)]
\endaligned
\end{eqnarray}
where we have used the fact that  $\Pi_C(\hat{x})\hat{U}_k^{\dag}(\hat{x})\Pi_C(\hat{x})=\Pi_C(\hat{x})\hat{U}_k(\hat{x})\Pi_C(\hat{x})=0$, this is because $\hat{U}_k(\hat{x})\Pi_C(\hat{x})\propto M_k(\hat{x})$, thus $\Pi_C(\hat{x})\hat{U}_k(\hat{x})\Pi_C(\hat{x})\propto \Pi_C(\hat{x})M_k(\hat{x})=\Pi_C(\hat{x})\Pi_E(\hat{x})E_k(\hat{x})\Pi_C(\hat{x})=0$ ( as $\Pi_C(\hat{x})\Pi_E(\hat{x})=0$).

 From the difference between $\rho_C(t+dt)$ and $\rho_C(t)$, we can then obtain the dynamics for the corrected state, which is
\begin{eqnarray}
\label{eq:eff_dyn}
\aligned
\frac{d\rho_C}{dt}=&L_x^C(\rho_C)\\
=&-i[H^C(x),\rho_C]+\sum_{k}[E_{k}^C(x)\rho E_{k}^{C\dag}(x)\\
&- \frac{1}{2}\{\Pi_CE_{k}^{\dag}(x)E_{k}(x)\Pi_C,\rho_C\}]\\
&+\sum_{k,j}\tilde{E}_{kj}^C(x)\rho_C \tilde{E}_{kj}^{C\dag}(x),
\endaligned
\end{eqnarray}
here 
\begin{eqnarray}
\label{eq:operators}
\aligned
& H^C(x) = \Pi_C(\hat{x}) H(x) \Pi_C(\hat{x}), \\
& E_{k}^C(x) = \Pi_C(\hat{x})E_{k}(x)\Pi_C(\hat{x}), \\
& \tilde{E}_{kj}^C(x) = \Pi_C(\hat{x})\hat{U}_{k}^{\dag}(\hat{x})E_{j}(x)\Pi_C(\hat{x}).
\endaligned
\end{eqnarray}

The operator, $L_x^C$, depends on $x$, which can be expanded around $\hat{x}$ up to the second order of $dx=x-\hat{x}$ as
\begin{eqnarray}
\aligned
L_x^C(\rho)=L_{\hat{x}}(\rho) + L_1(\rho)(x-\hat{x})+L_2(\rho)(x-\hat{x})^2,
\endaligned
\end{eqnarray}
where $L_1=\frac{\partial L_x^C}{\partial x}|_{x=\hat{x}}$ and $L_2=\frac{1}{2}\frac{\partial^2 L_x^C}{\partial^2 x}|_{x=\hat{x}}$.

We first expand each operator appeared in Eq.(\ref{eq:eff_dyn}) to the second order around $\hat{x}$, which are given by (we will use the over-dot to denote the derivatives with respect to $x$ which is then evaluated at the point $x=\hat{x}$, for example, $\dot{H}=\frac{\partial_x H(x)}{\partial x}|_{x=\hat{x}}$)

\begin{widetext}
\begin{eqnarray}
\aligned
H^C(x) =& \Pi_C(\hat{x}) H(x) \Pi_C(\hat{x})\\
=&\Pi_C(\hat{x}) H(\hat{x}) \Pi_C(\hat{x})+\Pi_C(\hat{x})\frac{\partial_xH(x)}{\partial x}|_{x=\hat{x}} \Pi_C(\hat{x})dx+\frac{1}{2}\Pi_C(\hat{x})\frac{\partial^2_xH(x)}{\partial x^2}|_{x=\hat{x}}\Pi_C(\hat{x})dx^2\\
=&\Pi_C(\hat{x}) H(\hat{x}) \Pi_C(\hat{x})+\Pi_C(\hat{x})\dot{H}\Pi_C(\hat{x})dx+\frac{1}{2}\Pi_C(\hat{x})\ddot{H}\Pi_C(\hat{x})dx^2,
\endaligned
\end{eqnarray}

\begin{eqnarray}
\aligned
E_{k}^C(x) =& \Pi_C(\hat{x})E_{k}(x)\Pi_C(\hat{x})\\
=&\Pi_C(\hat{x})E_{k}(\hat{x})\Pi_C(\hat{x})+\Pi_C(\hat{x})\frac{\partial_xE_k(x)}{\partial x}|_{x=\hat{x}} \Pi_C(\hat{x})dx+\frac{1}{2}\Pi_C(\hat{x})\frac{\partial^2_xE_k(x)}{\partial x^2}|_{x=\hat{x}}\Pi_C(\hat{x})dx^2\\
=&\alpha_k\Pi_C(\hat{x})+\Pi_C(\hat{x})\frac{\partial_xE_k(x)}{\partial x}|_{x=\hat{x}} \Pi_C(\hat{x})dx+\frac{1}{2}\Pi_C(\hat{x})\frac{\partial^2_xE_k(x)}{\partial x^2}|_{x=\hat{x}}\Pi_C(\hat{x})dx^2\\
=&\alpha_k\Pi_C(\hat{x})+\Pi_C(\hat{x})\dot{E}_k\Pi_C(\hat{x})dx+\frac{1}{2}\Pi_C(\hat{x})\ddot{E}_k\Pi_C(\hat{x})dx^2,\\
\endaligned
\end{eqnarray}

\begin{eqnarray}
\aligned
\tilde{E}_{kj}^C(x) =& \Pi_C(\hat{x})\hat{U}_{k}^{\dag}(\hat{x})E_{j}(x)\Pi_C(\hat{x})\\
=&\Pi_C(\hat{x})\hat{U}_{k}^{\dag}(\hat{x})E_{j}(\hat{x})\Pi_C(\hat{x})+\Pi_C(\hat{x})\hat{U}_{k}^{\dag}(\hat{x})\frac{\partial_xE_j(x)}{\partial x}|_{x=\hat{x}}\Pi_C(\hat{x})dx+\frac{1}{2}\Pi_C(\hat{x})\hat{U}_{k}^{\dag}(\hat{x})\frac{\partial_x^2E_j(x)}{\partial x^2}|_{x=\hat{x}}\Pi_C(\hat{x})dx^2\\
=&\delta_{kj}\Pi_C(\hat{x})+\Pi_C(\hat{x})\hat{U}_{k}^{\dag}(\hat{x})\frac{\partial_xE_j(x)}{\partial x}|_{x=\hat{x}}\Pi_C(\hat{x})dx+\frac{1}{2}\Pi_C(\hat{x})\hat{U}_{k}^{\dag}(\hat{x})\frac{\partial_x^2E_j(x)}{\partial x^2}|_{x=\hat{x}}\Pi_C(\hat{x})dx^2\\
=&\delta_{kj}\Pi_C(\hat{x})+\Pi_C(\hat{x})\hat{U}_{k}^{\dag}(\hat{x})\dot{E}_j\Pi_C(\hat{x})dx+\frac{1}{2}\Pi_C(\hat{x})\hat{U}_{k}^{\dag}(\hat{x})\ddot{E}_j\Pi_C(\hat{x})dx^2,\\
\endaligned
\end{eqnarray}

\begin{eqnarray}
\aligned
\Pi_C(\hat{x})&E_k^\dagger(x)E_k(x)\Pi_C(\hat{x})\\
=&\Pi_C(\hat{x})E_k^\dagger(\hat{x})E_k(\hat{x})\Pi_C(\hat{x})+\Pi_C(\hat{x})\frac{\partial_x [E_k^\dagger(x)E_k(x)]}{\partial x}|_{x=\hat{x}}\Pi_C(\hat{x})dx+\frac{1}{2}\Pi_C(\hat{x})\frac{\partial_x^2 [E_k^\dagger(x)E_k(x)]}{\partial x^2}|_{x=\hat{x}}\Pi_C(\hat{x})dx^2\\
=&\beta_{kk}\Pi_C(\hat{x})+\Pi_C(\hat{x})\frac{\partial_x [E_k^\dagger(x)E_k(x)]}{\partial x}|_{x=\hat{x}}\Pi_C(\hat{x})dx+\frac{1}{2}\Pi_C(\hat{x})\frac{\partial_x^2 [E_k^\dagger(x)E_k(x)]}{\partial x^2}|_{x=\hat{x}}\Pi_C(\hat{x})dx^2\\
=&\beta_{kk}\Pi_C(\hat{x})+\Pi_C(\hat{x})[\dot{E}_k^\dagger E_k(\hat{x})+E_k^\dagger(\hat{x})\dot{E}_k]\Pi_C(\hat{x})dx+\frac{1}{2}\Pi_C(\hat{x})[\ddot{E}_k^\dagger E_k(\hat{x})+2\dot{E}_k^\dagger\dot{E}_k+E_k^\dagger(\hat{x})\ddot{E}_k]\Pi_C(\hat{x})dx^2,
\endaligned
\end{eqnarray}

By substituting these expansions into Eq.(\ref{eq:eff_dyn}), we then get
\begin{eqnarray}
\aligned
L_{\hat{x}}(\rho)=&-i[\Pi_C(\hat{x}) H(\hat{x}) \Pi_C(\hat{x}),\rho],\\
L_1(\rho)=&-i[\Pi_C(\dot{H}+\frac{i}{2}\sum_kE_k^\dagger\dot{E}_k-\dot{E}_k^\dagger E_k)\Pi_C, \rho],\\
L_2(\rho)=&-i[\frac{1}{2}\Pi_C(\ddot{H}+\sum_k E_k^\dagger\ddot{E}_k-\ddot{E}_k^\dagger E_k)\Pi_C,\rho]\\
&+\sum_k[\Pi_C\dot{E}_{k}\Pi_C\rho\Pi_C\dot{E}_{k}^\dagger\Pi_C-\frac{1}{2}\{\Pi_C\dot{E}_{k}^{\dag}\dot{E}_k\Pi_C,\rho\}]\\
&+\sum_{k,j}\frac{1}{d_{kk}}\Pi_C(E_k^\dagger\dot{E}_j-\alpha_k^*\dot{E}_j)\Pi_C\rho\Pi_C(\dot{E}_j^\dagger E_k-\alpha_k\dot{E}_j^\dagger)\Pi_C.\\
\endaligned
\end{eqnarray}
\end{widetext}

%
\bibliography{sample}

\begin{thebibliography}{10}

\bibitem{giovannetti2011advances}
Vittorio Giovannetti, Seth Lloyd, and Lorenzo Maccone.
\newblock Advances in quantum metrology.
\newblock {\em Nature photonics}, 5(4):222--229, 2011.

\bibitem{giovannetti2006quantum}
Vittorio Giovannetti, Seth Lloyd, and Lorenzo Maccone.
\newblock Quantum metrology.
\newblock {\em Physical review letters}, 96(1):010401, 2006.

\bibitem{anisimov2010quantum}
Petr~M Anisimov, Gretchen~M Raterman, Aravind Chiruvelli, William~N Plick,
  Sean~D Huver, Hwang Lee, and Jonathan~P Dowling.
\newblock Quantum metrology with two-mode squeezed vacuum: parity detection
  beats the heisenberg limit.
\newblock {\em Physical review letters}, 104(10):103602, 2010.

\bibitem{braunstein1996generalized}
Samuel~L Braunstein, Carlton~M Caves, and Gerard~J Milburn.
\newblock Generalized uncertainty relations: theory, examples, and lorentz
  invariance.
\newblock {\em annals of physics}, 247(1):135--173, 1996.

\bibitem{paris2009quantum}
MATTEO G.~A. PARIS.
\newblock Quantum estimation for quantum technology.
\newblock {\em International Journal of Quantum Information},
  07(supp01):125--137, 2009.

\bibitem{Fujiwara2008}
A.~Fujiwara and H.~Imai.
\newblock A fibre bundle over manifolds of quantum channels and its application
  to quantum statistics.
\newblock {\em Journal of Physics A: Mathematical and Theoretical}, 41:255304,
  2008.

\bibitem{escher2012general}
B.~M. Escher, R.~L. de~Matos~Filho, and L.~Davidovich.
\newblock General framework for estimating the ultimate precision limit in
  noisy quantum-enhanced metrology.
\newblock {\em Nature Physics}, 7:406--411, 2011.

\bibitem{demkowicz2014using}
Rafal Demkowicz-Dobrzanski and Lorenzo Maccone.
\newblock Using entanglement against noise in quantum metrology.
\newblock {\em Physical review letters}, 113(25):250801, 2014.

\bibitem{demkowicz2012elusive}
R.~Demkowicz-Dobrzanski, J.~Kolodynski, and M.~Guta.
\newblock The elusive heisenberg limit in quantum-enhanced metrology.
\newblock {\em Nature Communications}, 3:1063, 2012.

\bibitem{schnabel2010quantum}
Roman Schnabel, Nergis Mavalvala, David~E McClelland, and Ping~K Lam.
\newblock Quantum metrology for gravitational wave astronomy.
\newblock {\em Nature communications}, 1:121, 2010.

\bibitem{huelga1997improvement}
Susanna~F Huelga, Chiara Macchiavello, Thomas Pellizzari, Artur~K Ekert,
  Martin~B Plenio, and J~Ignacio Cirac.
\newblock Improvement of frequency standards with quantum entanglement.
\newblock {\em Physical Review Letters}, 79(20):3865, 1997.

\bibitem{chin2012quantum}
Alex~W Chin, Susana~F Huelga, and Martin~B Plenio.
\newblock Quantum metrology in non-markovian environments.
\newblock {\em Physical review letters}, 109(23):233601, 2012.

\bibitem{HallPRX}
Michael J.~W. Hall and Howard~M. Wiseman.
\newblock Does nonlinear metrology offer improved resolution? answers from
  quantum information theory.
\newblock {\em Phys. Rev. X}, 2:041006, Oct 2012.

\bibitem{Berry2015}
Dominic~W. Berry, Mankei Tsang, Michael J.~W. Hall, and Howard~M. Wiseman.
\newblock Quantum bell-ziv-zakai bounds and heisenberg limits for waveform
  estimation.
\newblock {\em Phys. Rev. X}, 5:031018, Aug 2015.

\bibitem{Alipour2014}
S.~Alipour, M.~Mehboudi, and A.~T. Rezakhani.
\newblock Quantum metrology in open systems: Dissipative cram\'er-rao bound.
\newblock {\em Phys. Rev. Lett.}, 112:120405, Mar 2014.

\bibitem{Beau2017}
M.~Beau and A.~del Campo.
\newblock Nonlinear quantum metrology of many-body open systems.
\newblock {\em Phys. Rev. Lett.}, 119:010403, Jul 2017.

\bibitem{Liu_2019}
Jing Liu, Haidong Yuan, Xiao-Ming Lu, and Xiaoguang Wang.
\newblock Quantum fisher information matrix and multiparameter estimation.
\newblock {\em Journal of Physics A: Mathematical and Theoretical},
  53(2):023001, dec 2019.

\bibitem{ligo2011gravitational}
LIGO.
\newblock A gravitational wave observatory operating beyond the quantum
  shot-noise limit: Squeezed light in application.
\newblock {\em Nature Physics}, 7:962--965, 2011.

\bibitem{joo2011quantum}
Jaewoo Joo, William~J Munro, and Timothy~P Spiller.
\newblock Quantum metrology with entangled coherent states.
\newblock {\em Physical review letters}, 107(8):083601, 2011.

\bibitem{higgins2007entanglement}
Brendon~L Higgins, Dominic~W Berry, Stephen~D Bartlett, Howard~M Wiseman, and
  Geoff~J Pryde.
\newblock Entanglement-free heisenberg-limited phase estimation.
\newblock {\em Nature}, 450:393--396, 2007.

\bibitem{kolobov1999spatial}
Mikhail~I Kolobov.
\newblock The spatial behavior of nonclassical light.
\newblock {\em Reviews of Modern Physics}, 71(5):1539, 1999.

\bibitem{lugiato2002quantum}
LA~Lugiato, A~Gatti, and E~Brambilla.
\newblock Quantum imaging.
\newblock {\em Journal of Optics B: Quantum and semiclassical optics},
  4(3):S176, 2002.

\bibitem{morris2015imaging}
Peter~A Morris, Reuben~S Aspden, Jessica~EC Bell, Robert~W Boyd, and Miles~J
  Padgett.
\newblock Imaging with a small number of photons.
\newblock {\em Nature communications}, 6:5913, 2015.

\bibitem{roga2016security}
Wojciech Roga and John Jeffers.
\newblock Security against jamming and noise exclusion in imaging.
\newblock {\em Physical Review A}, 94(3):032301, 2016.

\bibitem{tsang2016quantum}
Mankei Tsang, Ranjith Nair, and Xiao-Ming Lu.
\newblock Quantum theory of superresolution for two incoherent optical point
  sources.
\newblock {\em Physical Review X}, 6(3):031033, 2016.

\bibitem{shapiro2009quantum}
Jeffrey~H Shapiro and Seth Lloyd.
\newblock Quantum illumination versus coherent-state target detection.
\newblock {\em New Journal of Physics}, 11(6):063045, 2009.

\bibitem{lopaeva2013experimental}
ED~Lopaeva, I~Ruo Berchera, IP~Degiovanni, S~Olivares, G~Brida, and M~Genovese.
\newblock Experimental realization of quantum illumination.
\newblock {\em Physical review letters}, 110(15):153603, 2013.

\bibitem{dowling1998correlated}
Jonathan~P Dowling.
\newblock Correlated input-port, matter-wave interferometer: Quantum-noise
  limits to the atom-laser gyroscope.
\newblock {\em Physical Review A}, 57(6):4736, 1998.

\bibitem{bollinger1996optimal}
JJ~Bollinger, Wayne~M Itano, DJ~Wineland, and DJ~Heinzen.
\newblock Optimal frequency measurements with maximally correlated states.
\newblock {\em Physical Review A}, 54(6):R4649, 1996.

\bibitem{buvzek1999optimal}
Vladim{\'\i}r Bu{\v{z}}ek, Radoslav Derka, and Serge Massar.
\newblock Optimal quantum clocks.
\newblock {\em Physical review letters}, 82(10):2207, 1999.

\bibitem{leibfried2004toward}
D~Leibfried, Murray~D Barrett, T~Schaetz, J~Britton, J~Chiaverini, Wayne~M
  Itano, John~D Jost, Christopher Langer, and David~J Wineland.
\newblock Toward heisenberg-limited spectroscopy with multiparticle entangled
  states.
\newblock {\em Science}, 304(5676):1476--1478, 2004.

\bibitem{roos2007designer}
CF~Roos, M~Chwalla, K~Kim, M~Riebe, and R~Blatt.
\newblock 'designer atoms' for quantum metrology.
\newblock {\em Nature}, 443:316, 2006.

\bibitem{derevianko2011colloquium}
Andrei Derevianko and Hidetoshi Katori.
\newblock Colloquium: Physics of optical lattice clocks.
\newblock {\em Reviews of Modern Physics}, 83(2):331, 2011.

\bibitem{ludlow2015optical}
Andrew~D Ludlow, Martin~M Boyd, Jun Ye, Ekkehard Peik, and Piet~O Schmidt.
\newblock Optical atomic clocks.
\newblock {\em Reviews of Modern Physics}, 87(2):637, 2015.

\bibitem{borregaard2013near}
Johannes Borregaard and Anders~S S{\o}rensen.
\newblock Near-heisenberg-limited atomic clocks in the presence of decoherence.
\newblock {\em Physical review letters}, 111(9):090801, 2013.

\bibitem{yuanfd}
Haidong Yuan and Chi-Hang~Fred Fung.
\newblock Fidelity and fisher information on quantum channels.
\newblock {\em New Journal of Physics}, 19(11):113039, 2017.

\bibitem{Plenio2000}
C.~Macchiavello, S.~F. Huelga, J.~I. Cirac, A.~K. Ekert, and M.~B. Plenio.
\newblock {\em Decoherence and quantum error correction in frequency
  standards}.
\newblock Kluwer Academic/Plenum Publishers, New York, 2000.

\bibitem{Dur2014}
W.~D\"ur, M.~Skotiniotis, F.~Fr\"owis, and B.~Kraus.
\newblock Improved quantum metrology using quantum error correction.
\newblock {\em Phys. Rev. Lett.}, 112:080801, 2014.

\bibitem{Arrad2014}
G.~Arrad, Y.~Vinkler, D.~Aharonov, and A.~Retzker.
\newblock Increasing sensing resolution with error correction.
\newblock {\em Phys. Rev. Lett.}, 112:150801, 2014.

\bibitem{Kessler2014}
E.~M. Kessler, I.~Lovchinsky, A.~O. Sushkov, and M.~D. Lukin.
\newblock Quantum error correction for metrology.
\newblock {\em Phys. Rev. Lett.}, 112:150802, 2014.

\bibitem{Ozeri2013}
R.~Ozeri.
\newblock Heisenberg limited metrology using quantum error-correction codes.
\newblock {\em arXiv}, page 1310.3432, 2013.

\bibitem{Unden2016}
Thomas Unden, Priya Balasubramanian, Daniel Louzon, Yuval Vinkler, Martin~B.
  Plenio, Matthew Markham, Daniel Twitchen, Alastair Stacey, Igor Lovchinsky,
  Alexander~O. Sushkov, Mikhail~D. Lukin, Alex Retzker, Boris Naydenov, Liam~P.
  McGuinness, and Fedor Jelezko.
\newblock Quantum metrology enhanced by repetitive quantum error correction.
\newblock {\em Phys. Rev. Lett.}, 116:230502, Jun 2016.

\bibitem{Sekatski2017}
Pavel Sekatski, Michalis Skotiniotis, Janek Kolodynski, and Wolfgang Dur.
\newblock Quantum metrology with full and fast quantum control.
\newblock {\em {Quantum}}, 1:27, September 2017.

\bibitem{Rafal2017}
Rafal Demkowicz-Dobrzanski, Jan Czajkowski, and Pavel Sekatski.
\newblock Adaptive quantum metrology under general markovian noise.
\newblock {\em Phys. Rev. X}, 7:041009, Oct 2017.

\bibitem{Zhou2018}
Sisi Zhou, Mengzhen Zhang, John Preskill, and Liang Jiang.
\newblock Achieving the heisenberg limit in quantum metrology using quantum
  error correction.
\newblock {\em Nature Communications}, 9(1):78, 2018.

\bibitem{Layden2018}
David Layden and Paola Cappellaro.
\newblock Spatial noise filtering through error correction for quantum sensing.
\newblock {\em npj Quantum Information}, 4:30, Jul 2018.

\bibitem{Layden2019}
David Layden, Sisi Zhou, Paola Cappellaro, and Liang Jiang.
\newblock Ancilla-free quantum error correction codes for quantum metrology.
\newblock {\em Phys. Rev. Lett.}, 122:040502, Jan 2019.

\bibitem{Zhou2019}
Sisi Zhou and Liang Jiang.
\newblock Optimal approximate quantum error correction for quantum metrology.
\newblock {\em S Zhou, L Jiang - arXiv preprint arXiv:1910.08472}, 2019.

\bibitem{Layden2020}
David Layden, Louisa~Ruixue Huang, and Paola Cappellaro.
\newblock Robustness-optimized quantum error correction.
\newblock {\em Quantum Science and Technology}, 2020.

\bibitem{Holevo}
AS~Holevo.
\newblock {\em Probabilistic and Quantum Aspects of Quantum Theory}.
\newblock North-Holland, Amsterdam, 1982.

\bibitem{helstrom1976quantum}
Carl~Wilhelm Helstrom.
\newblock {\em Quantum detection and estimation theory}.
\newblock Academic press, 1976.

\bibitem{braunstein1994statistical}
Samuel~L Braunstein and Carlton~M Caves.
\newblock Statistical distance and the geometry of quantum states.
\newblock {\em Physical Review Letters}, 72(22):3439, 1994.

\bibitem{Wilcox1967}
R.~M. Wilcox.
\newblock Exponential operators and parameter differentiation in quantum
  physics.
\newblock {\em Journal of Mathematical Physics}, 8(4):962--982, 1967.

\bibitem{Brody_2013}
Dorje Brody and Eva-Maria Graefe.
\newblock Information geometry of complex hamiltonians and exceptional points.
\newblock {\em Entropy}, 15(12):3361--3378, Aug 2013.

\bibitem{Pang2014PRA}
Shengshi Pang and Todd~A. Brun.
\newblock Quantum metrology for a general hamiltonian parameter.
\newblock {\em Phys. Rev. A}, 90:022117, Aug 2014.

\bibitem{Liu2015}
Jing Liu, Xiao-Xing Jing, and Xiaoguang Wang.
\newblock Quantum metrology with unitary parametrization processes.
\newblock {\em Scientific Reports}, 5:8565, 2015.

\bibitem{Jing2015}
Xiao-Xing Jing, Jing Liu, Heng-Na Xiong, and Xiaoguang Wang.
\newblock Maximal quantum fisher information for general su(2) parametrization
  processes.
\newblock {\em Phys. Rev. A}, 92:012312, Jul 2015.

\bibitem{yuan2015optimal}
Haidong Yuan and Chi-Hang~Fred Fung.
\newblock Optimal feedback scheme and universal time scaling for hamiltonian
  parameter estimation.
\newblock {\em Physical review letters}, 115(11):110401, 2015.

\bibitem{Wiebe2014}
Nathan Wiebe, Christopher Granade, Christopher Ferrie, and D.~G. Cory.
\newblock Hamiltonian learning and certification using quantum resources.
\newblock {\em Phys. Rev. Lett.}, 112:190501, May 2014.

\bibitem{Sinitsyn_2016}
Nikolai~A Sinitsyn and Yuriy~V Pershin.
\newblock The theory of spin noise spectroscopy: a review.
\newblock {\em Reports on Progress in Physics}, 79(10):106501, sep 2016.

\bibitem{Norris2016}
Leigh~M. Norris, Gerardo~A. Paz-Silva, and Lorenza Viola.
\newblock Qubit noise spectroscopy for non-gaussian dephasing environments.
\newblock {\em Phys. Rev. Lett.}, 116:150503, Apr 2016.

\bibitem{Goldstein2011}
G.~Goldstein, P.~Cappellaro, J.~R. Maze, J.~S. Hodges, L.~Jiang, A.~S.
  S\o{}rensen, and M.~D. Lukin.
\newblock Environment-assisted precision measurement.
\newblock {\em Phys. Rev. Lett.}, 106:140502, Apr 2011.

\bibitem{Cappellaro2012}
P.~Cappellaro, G.~Goldstein, J.~S. Hodges, L.~Jiang, J.~R. Maze, A.~S.
  S\o{}rensen, and M.~D. Lukin.
\newblock Environment-assisted metrology with spin qubits.
\newblock {\em Phys. Rev. A}, 85:032336, Mar 2012.


\bibitem{HUBNER1992239}
Matthias Hübner.
\newblock Explicit computation of the bures distance for density matrices.
\newblock {\em Physics Letters A}, 163(4):239 -- 242, 1992.

\bibitem{note2}
The previous argument is heuristic as the dynamics is not exactly unitary.

\bibitem{Pang2017}
S.~Pang and A.~N. Jordan.
\newblock Optimal adaptive control for quantum metrology with time-dependent
  hamiltonians.
\newblock {\em Nature Communications}, 8:14695, 2017.

\bibitem{Schmitt832}
Simon Schmitt, Tuvia Gefen, Felix~M. St{\"u}rner, Thomas Unden, Gerhard Wolff,
  Christoph M{\"u}ller, Jochen Scheuer, Boris Naydenov, Matthew Markham,
  Sebastien Pezzagna, Jan Meijer, Ilai Schwarz, Martin Plenio, Alex Retzker,
  Liam~P. McGuinness, and Fedor Jelezko.
\newblock Submillihertz magnetic spectroscopy performed with a nanoscale
  quantum sensor.
\newblock {\em Science}, 356(6340):832--837, 2017.

\bibitem{Boss837}
J.~M. Boss, K.~S. Cujia, J.~Zopes, and C.~L. Degen.
\newblock Quantum sensing with arbitrary frequency resolution.
\newblock {\em Science}, 356(6340):837--840, 2017.

\bibitem{Naghiloo2017}
M.~Naghiloo, A.~N. Jordan, and K.~W. Murch.
\newblock Achieving optimal quantum acceleration of frequency estimation using
  adaptive coherent control.
\newblock {\em Phys. Rev. Lett.}, 119:180801, Nov 2017.

\bibitem{Glenn2018}
David~R. Glenn, Dominik~B. Bucher, Junghyun Lee, Mikhail~D. Lukin, Hongkun
  Park, and Ronald~L. Walsworth.
\newblock High-resolution magnetic resonance spectroscopy using a solid-state
  spin sensor.
\newblock {\em Nature}, 555:351--354, 2018.

\bibitem{Note1}
https://github.com/anschen1994/CorEnhance.

\bibitem{Braun2017}
Julien Mathieu~Elias Fra\"{\i}sse and Daniel Braun.
\newblock Enhancing sensitivity in quantum metrology by hamiltonian extensions.
\newblock {\em Phys. Rev. A}, 95:062342, Jun 2017.

\end{thebibliography}
\bibliographystyle{unsrt}

\end{document}